\documentclass[aps, pre, a4paper, floatfix,  twocolumn, longbibliography, nofootinbib]{revtex4-1}

\usepackage[utf8]{inputenc}
\usepackage{epsfig}
\usepackage[T1]{fontenc}
\usepackage[english]{babel}
\usepackage[table]{xcolor}
\usepackage{t1enc}
\usepackage{graphicx}
\usepackage{amssymb}
\usepackage{amsmath}
\usepackage{relsize}
\usepackage{overpic}
\usepackage{bm}
\usepackage[section]{placeins}

\usepackage[normalem]{ulem}

\newcommand{\avg}[1]{{\left<#1\right>}}

\newcommand{\ceil}[1]{{\lceil #1\rceil}}

\newcommand{\dd}{\mathrm{d}}

\usepackage{mathtools}
\def\multiset#1#2{\ensuremath{\left(\kern-.3em\left(\genfrac{}{}{0pt}{}{#1}{#2}\right)\kern-.3em\right)}}

\usepackage{amsmath}

\newcolumntype{C}{>{\centering\arraybackslash}m{.32\textwidth}}

\usepackage{verbatim}
\usepackage{overpic}
\usepackage{booktabs}
\usepackage{placeins}

\begin{document}

\title{Nonparametric Bayesian inference of the microcanonical stochastic block model}

\author{Tiago P. Peixoto}
\email{t.peixoto@bath.ac.uk}
\affiliation{Department of Mathematical Sciences and Centre for Networks
and Collective Behaviour, University of Bath, Claverton Down, Bath BA2
7AY, United Kingdom}
\affiliation{ISI Foundation, Via Alassio 11/c, 10126 Torino, Italy}

\pacs{89.75.Hc 02.50.Tt 89.70.Cf}

\begin{abstract}

A principled approach to characterize the hidden structure of networks
is to formulate generative models, and then infer their parameters from
data. When the desired structure is composed of modules or
"communities", a suitable choice for this task is the stochastic block
model (SBM), where nodes are divided into groups, and the placement of
edges is conditioned on the group memberships. Here, we present a
nonparametric Bayesian method to infer the modular structure of
empirical networks, including the number of modules and their
hierarchical organization. We focus on a microcanonical variant of the
SBM, where the structure is imposed via hard constraints, i.e. the
generated networks are not allowed to violate the patterns imposed by
the model. We show how this simple model variation allows simultaneously
for two important improvements over more traditional inference
approaches: 1. Deeper Bayesian hierarchies, with noninformative priors
replaced by sequences of priors and hyperpriors, that not only remove
limitations that seriously degrade the inference on large networks, but
also reveal structures at multiple scales; 2. A very efficient inference
algorithm that scales well not only for networks with a large number of
nodes and edges, but also with an unlimited number of modules. We show
also how this approach can be used to sample modular hierarchies from
the posterior distribution, as well as to perform model selection. We
discuss and analyze the differences between sampling from the posterior
and simply finding the single parameter estimate that maximizes
it. Furthermore, we expose a direct equivalence between our
microcanonical approach and alternative derivations based on the
canonical SBM.

\end{abstract}

\maketitle

\section{Introduction}

One of the most basic goals in the study of social, biological and
technological networks is the characterization of their structural
patterns. As these systems become large, this quickly becomes a
nontrivial problem, as naive methods of inspection are no longer useful,
and simple statistics often hide crucial information. A popular approach
to this problem is the development of methods that divide the network by
grouping together nodes that share similar features, thereby reducing it
to a more manageable size, and in the process revealing any latent
modular organization. This is the core idea behind a very large number
of heuristic methods proposed in the last decade and a
half~\cite{fortunato_community_2010,fortunato_community_2016}, which
despite sharing the same motivation differ substantially from each
other, due mainly to the various ways this intuitive idea can be
implemented concretely. Over time it has become clear that most of these
methods are marred by serious limitations, such as the incapacity of
distinguishing structure from noise~\cite{guimera_modularity_2004} and
to find small structures in large
systems~\cite{fortunato_resolution_2007}, as well as the fact that the
same method often yields multiple diverging results for the same
network~\cite{good_performance_2010}, and that the outcomes of most
methods agree neither with each other~\cite{fortunato_community_2016}
nor with known node annotations~\cite{hric_community_2014}.

Like some more recent works in this area, here we follow a different and
arguably more principled path, designed to overcome some of these
limitations. Namely, instead of formulating heuristics, we construct
probabilistic generative models of networks, that include the
aforementioned idea of modular structure as parameters to the model. The
modular organization is then determined by inferring these parameters
from data, using well-founded methods from Bayesian inference and
statistical physics. In this context, the problem of separating
structure from noise is dealt with by employing nonparametric inference,
where generative processes for the model parameters are also formulated
via prior probabilities. Additionally, the comparison of different modular
partitions --- obtained either from the same or from different models
incorporating potentially different ideas about modular organization
--- can be performed probabilistically, and amount to a comparison of
alternative generative hypotheses according to statistical evidence.

In this work, we focus on a specific family of generative models based
on the stochastic block model (SBM)~\cite{holland_stochastic_1983},
where nodes are divided into groups, and the edges are placed randomly
between nodes, with probabilities that depend on their group
memberships. In particular, we consider a \emph{microcanonical}
variation of this family, where the structural constraints are imposed
strictly across the ensemble, as opposed to only on average, as is more
typically done. We show how this approach makes it easier to incorporate
more elaborate generative models, where parameters are sampled from
conditioned prior probabilities, which themselves are sampled from
hyperprior distributions. This yields a more powerful method that
reveals the hierarchical organization of networks in multiple scales,
and has a much increased capacity of finding statistically significant
structures in large data. Furthermore, we show how this particular
formulation allows for a very efficient inference algorithm that scales
well not only for networks with a large number of nodes and edges, but
also with an unlimited number of modules --- in contrast to the majority
of other similar inference algorithms that become increasingly slower as
the number of groups becomes large.

The approach taken here builds upon ideas from previous
work~\cite{peixoto_parsimonious_2013, peixoto_efficient_2014,
peixoto_hierarchical_2014}, but here we focus on obtaining hierarchical
network partitions that are \emph{sampled} from the posterior
distribution, instead of finding only the most likely partition, which
requires a different ansatz. We also show how model selection can be
used to choose between different model variants according to the
statistical evidence available in the data, and how the method fares for
a variety of empirical networks. Furthermore we show that the
microcanonical formulation used here is --- in its most basic form ---
equivalent to a specific Bayesian formulation of the ``canonical'' SBM,
and thus we establish a bridge between both approaches.

The paper is divided as follows. We begin in Sec.~\ref{sec:dc-sbm} with
the microcanonical SBM, and follow in Sec.~\ref{sec:inference} with the
outline of the nonparametric inference approach, by describing in turn
the priors and hyperpriors of the different set of parameters. In
Sec.~\ref{sec:canonical} we show how the microcanonical formulation is
related to the more usual canonical approach, and in
Sec.~\ref{sec:resolution} we analyze the limitations of the inference
procedure, and we show how the hierarchical approach is capable of
finding a much larger number of groups in large networks. In
Sec.~\ref{sec:mcmc} we present an efficient MCMC algorithm to sample
hierarchical partitions from the posterior distribution. In
Sec.~\ref{sec:comparison} we show how different model variations can be
compared, and in Sec.~\ref{sec:empirical} we show how the same
variations behave for empirical networks. We finalize in
Sec.~\ref{sec:conclusion} with a discussion.

\section{The microcanonical degree-corrected SBM}\label{sec:dc-sbm}

We begin with a ``degree-corrected'' version of the
SBM~\cite{karrer_stochastic_2011} (DC-SBM), where in addition to the
modular structure, the networks generated possess a prescribed degree
sequence. However, differently from its original definition, here we
assume that the degree sequence is fixed exactly, instead of only in
expectation. We will see later that the non-degree-corrected version of
the model (NDC-SBM) can be obtained from this more general formulation
as a special case.

The parameters of the model are the partition $\bm{b}=\{b_i\}$ of $N$
nodes into $B$ groups, where $b_i\in [1, B]$ is the group membership of
node $i$, the degree sequence $\bm{k}=\{k_i\}$, and the matrix of edge
counts between groups $\bm{e}=\{e_{rs}\}$, where $e_{rs}$ is the number
of edges between groups $r$ and $s$ (for convenience of notation,
$e_{rr}$ is \emph{twice} the number of edges inside group $r$).  Given
these parameters, networks are generated like in the configuration
model~\cite{bollobas_probabilistic_1980,fosdick_configuring_2016}: To
each vertex $i$ is attributed $k_i$ half-edges (or ``stubs''), which are
paired randomly to each other
--- allowing for multiple pairings between the same two nodes as
well as self-loops --- respecting the constraint that between groups $r$
and $s$ there are exactly $e_{rs}$ pairings. Assuming momentarily that
the half-edges are distinguishable, the number of possible pairings that
satisfy this constraint is given by
\begin{equation}\label{eq:omega}
  \Omega(\bm{e}) = \frac{\prod_re_r!}{\prod_{r<s}e_{rs}!\prod_re_{rr}!!},
\end{equation}
where $e_r=\sum_se_{rs}$ and $(2m)!!=2^mm!$.  However, many different
pairings correspond to the same graph. Given an adjacency matrix
$\bm{A}$, the number of different half-edge pairings to which it
corresponds is analogously given by
\begin{equation}\label{eq:xi}
  \Xi(\bm{A}) = \frac{\prod_ik_i!}{\prod_{i<j}A_{ij}!\prod_iA_{ii}!!}.
\end{equation}
Hence, the probability of observing a particular network given the model
parameters is simply the ratio between these two numbers,
\begin{equation}\label{eq:model_dc}
  P(\bm{A}|\bm{k},\bm{e},\bm{b}) = \frac{\Xi(\bm{A})}{\Omega(\bm{e})}.
\end{equation}
(Naturally, the above likelihood only holds if the network $\bm{A}$
matches exactly the hard constraints imposed by the parameters,
i.e. $e_{rs} = \sum_{ij}A_{ij}\delta_{b_i,r}\delta_{b_j,s}$ and
$k_i=\sum_jA_{ij}$, otherwise the likelihood is zero. In order to leave
the expressions uncluttered, we will always implicitly assume that the
hard constraints must hold for the likelihoods to be nonzero.)

The model above generates graphs with multiple edges between nodes,
which may not be strictly appropriate for many types of networks where
this cannot occur. However --- as is true with the traditional
configuration model --- the probability of multiple edges will decrease
with $1/N$ for sparse networks with $E \propto N$ edges, and hence their
occurrence can be neglected as $N$ becomes large.

\section{Nonparametric Bayesian inference}\label{sec:inference}
\begin{figure}
  \begin{tabular}{cc}
    \begin{minipage}{.49\columnwidth}\smaller
    \includegraphics[width=\textwidth]{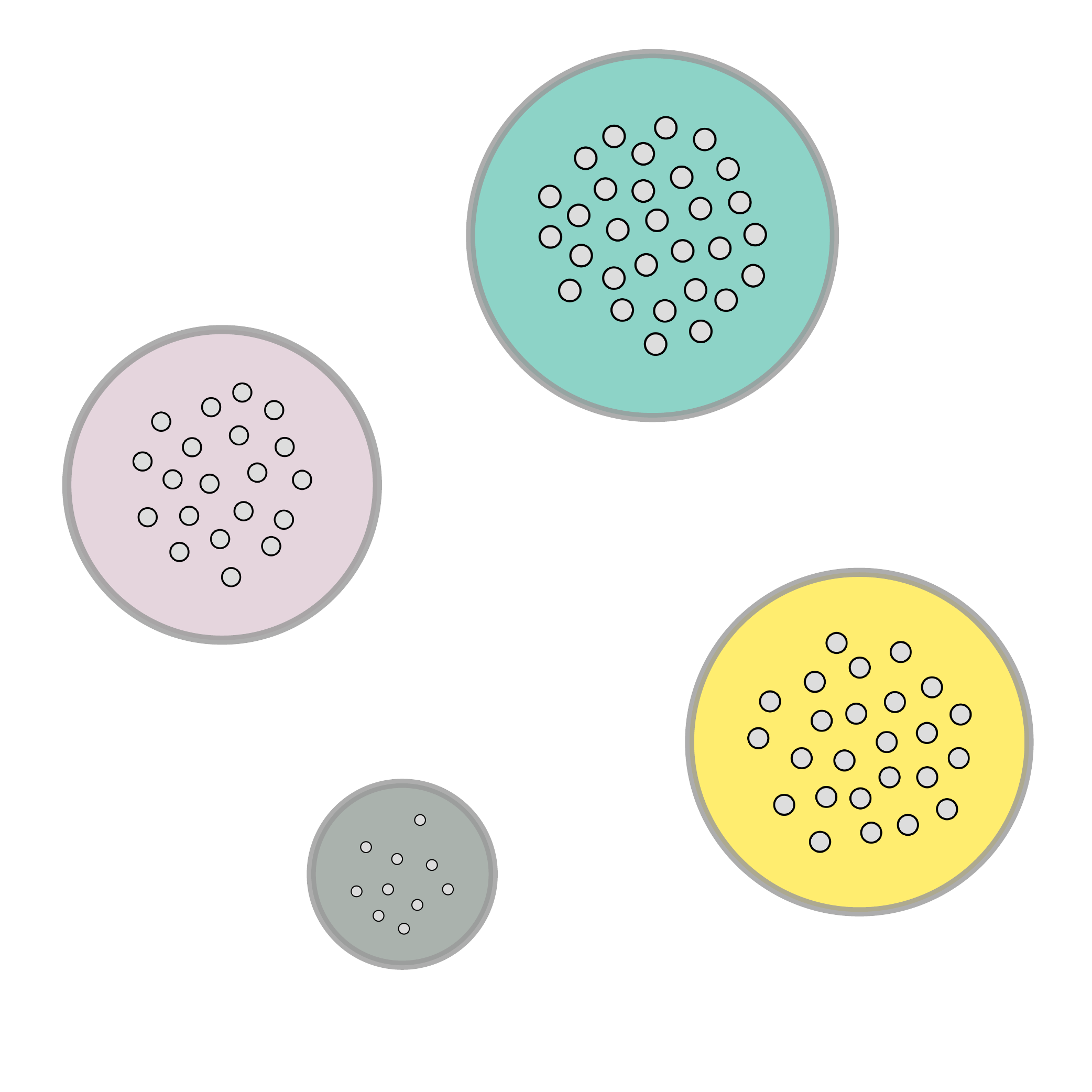}\\
    (a) Node partition, $P(\bm{b}).$
    \end{minipage}
    &
    \begin{minipage}{.49\columnwidth}\smaller
    \includegraphics[width=\textwidth]{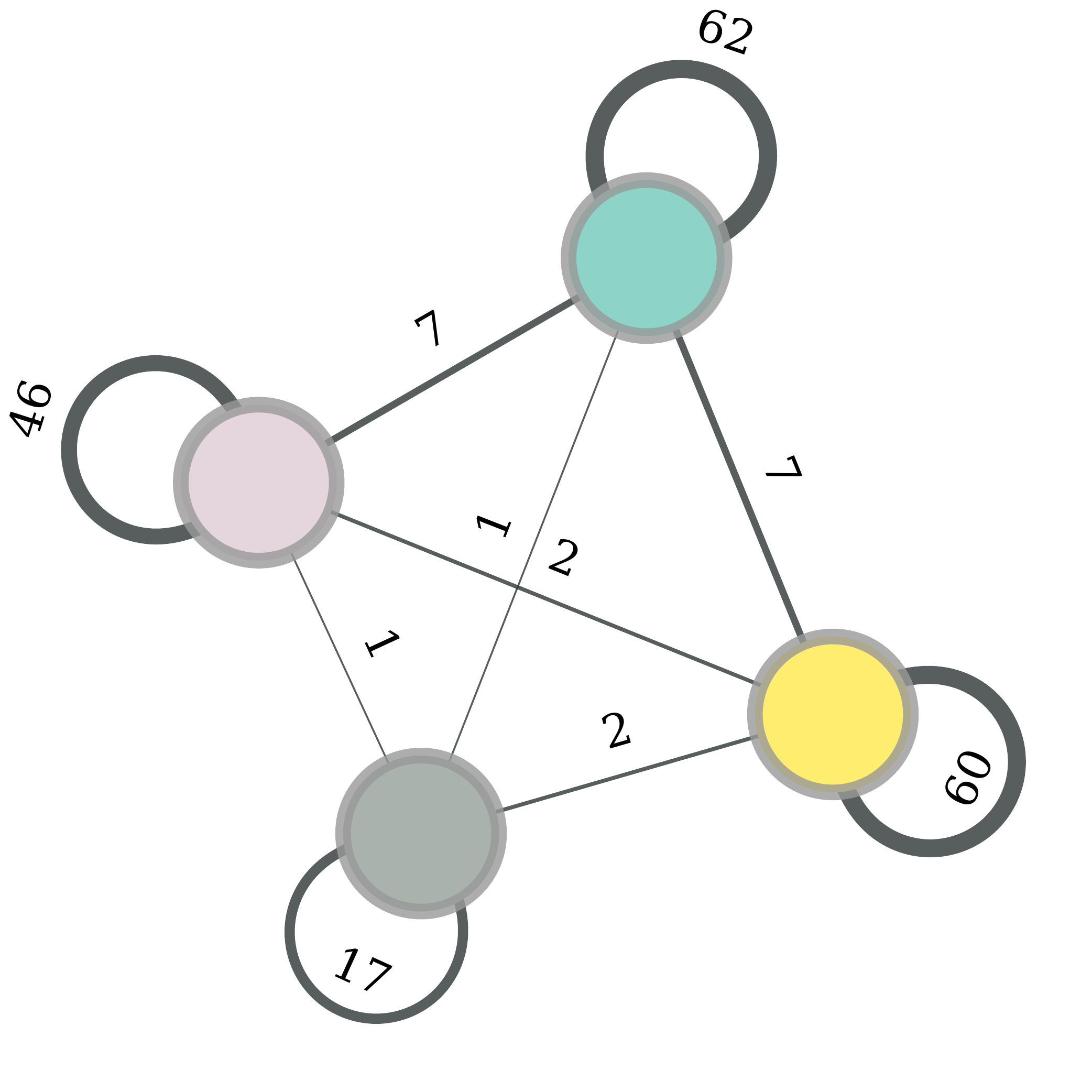}\\
    (b) Edge counts, $P(\bm{e}|\bm{b}).$
    \end{minipage}\\
    \begin{minipage}{.49\columnwidth}\smaller
    \includegraphics[width=\textwidth]{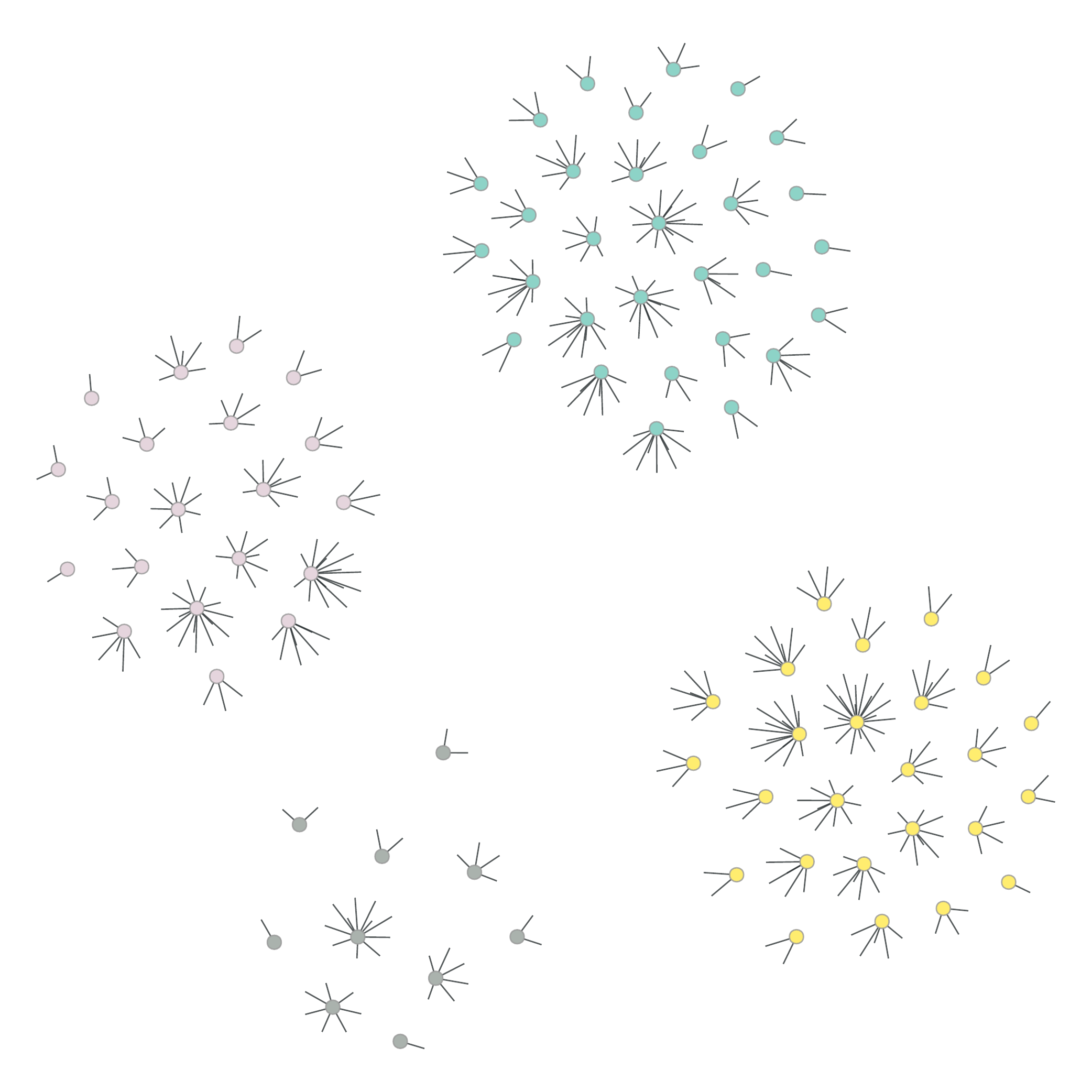}\\
    (c) Degrees, $P(\bm{k}|\bm{e},\bm{b}).$
    \end{minipage}
    &
    \begin{minipage}{.49\columnwidth}\smaller
    \includegraphics[width=\textwidth]{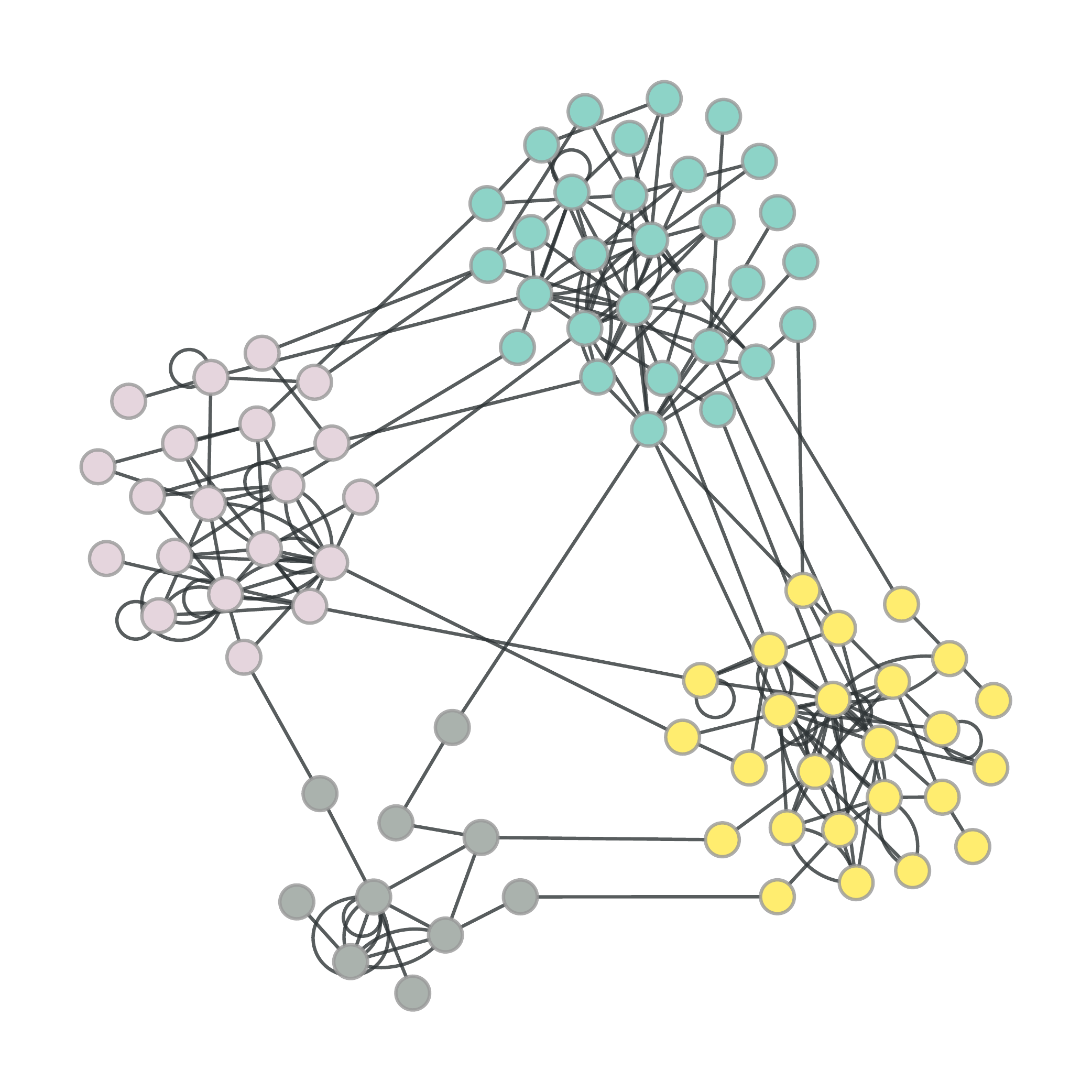}\\
    (d) Network, $P(\bm{A}|\bm{k},\bm{e},\bm{b}).$
    \end{minipage}\\
  \end{tabular}

  \caption{Illustration of the complete nonparametric generative process
  for the DC-SBM considered in this work. First the partition of the
  nodes is sampled (a), followed by the edge counts between groups (b),
  the degrees of the nodes (c) and finally the network itself
  (d).\label{fig:generative}}
\end{figure}

Although one could find the best divisions of the network by maximizing,
or sampling from Eq.~\ref{eq:model_dc} directly, this requires the
number of groups $B$ to be known in advance, i.e. it is a
\emph{parametric} inference procedure that requires certain properties
of the model to be determined \emph{a priori}. Instead, here we wish to
formulate a \emph{nonparametric} framework, where the number of groups
as well as any other model parameter is determined from the data
itself. In order to do this, we need to write the full joint
distribution for the data and the parameters,
\begin{equation}\label{eq:joint}
  P(\bm{A},\bm{k},\bm{e},\bm{b}) = P(\bm{A}|\bm{k},\bm{e},\bm{b})P(\bm{k}|\bm{e},\bm{b})P(\bm{e}|\bm{b})P(\bm{b}),
\end{equation}
where $P(\bm{k}|\bm{e},\bm{b})$, $P(\bm{e}|\bm{b})$, and $P(\bm{b})$ are
prior probabilities. The above defines a complete generative model for
the data and parameters, as illustrated in Fig.~\ref{fig:generative}.

Based on this, we can obtain the \emph{posterior}
distribution of network partitions,
\begin{equation}\label{eq:posterior}
  P(\bm{b} | \bm{A}) = \frac{P(\bm{A},\bm{b})}{P(\bm{A})},
\end{equation}
where the normalization constant
\begin{equation}\label{eq:evidence}
  P(\bm{A}) = \sum_{\bm{b}}P(\bm{A},\bm{b})
\end{equation}
is called the \emph{model evidence}, and $P(\bm{A},\bm{b})$ is the
marginal distribution corresponding to the joint probability summed over
the remaining parameters,
\begin{align}\label{eq:joint-marginal}
  P(\bm{A},\bm{b}) &= \sum_{\bm{k},\bm{e}}P(\bm{A},\bm{k},\bm{e},\bm{b})\\
  &= P(\bm{A},\hat{\bm{e}}(\bm{A},\bm{b}),\hat{\bm{k}}(\bm{A}),\bm{b}),
\end{align}
where $\hat{\bm{e}}$ and $\hat{\bm{k}}$ above are the only parameter
choices that fulfill the model constraints compatible with the
particular instance of the network $\bm{A}$ and the partition $\bm{b}$,
i.e.
\begin{align}
  \hat{e}_{rs}(\bm{A},\bm{b}) &= \sum_{ij}A_{ij}\delta_{b_i,r}\delta_{b_j,s}\label{eq:hat_e},\\
  \hat{k}_i(\bm{A}) &= \sum_{j}A_{ij}.\label{eq:hat_k}
\end{align}
In other words, any other choice $\bm{k}\ne\hat{\bm{k}}$ or
$\bm{e}\ne\hat{\bm{e}}$ inserted in Eq.~\ref{eq:model_dc} will result in
networks that are invariably different from the particular value of
$\bm{A}$ used in Eqs.~\ref{eq:joint} to~\ref{eq:hat_k}, and thus the
corresponding joint probability in Eq.~\ref{eq:joint-marginal} will be
zero. From this, we already observe a useful property of the
microcanonical formulation: Because of the hard constraints, there is no
difference between the joint and marginal probabilities. This means that
we encounter no additional computational difficulty in obtaining the
marginal probability after we have determined our priors. This is in
general different from ``canonical'' model formulations with continuous
parameters, where the marginal likelihood needs to be obtained via
integration, which sometimes cannot be done exactly, even if the choice
of prior happens to be well motivated. In the particular case of the
SBM, there are in fact typical canonical formulations where the marginal
likelihood can be computed
exactly~\cite{guimera_missing_2009,yan_active_2010,peixoto_hierarchical_2014,
come_model_2015,newman_estimating_2016}, but this has been done only for
simple non-informative or conjugate priors, which leads to serious
problems for large networks, as we discuss further in
Sec.~\ref{sec:resolution}. Here, instead, we can focus on priors that
are chosen according to more fundamental principles, without having to
worry about the computation of the marginal likelihood, provided the
priors themselves can be computed. As we will show below, this will
allow deeper Bayesian hierarchies to be developed, which make fewer
assumptions about the data generating process, and lifts important
practical limitations present in shallower approaches.

\subsection{Sampling vs. optimization and the minimum description length principle (MDL).}
\label{sec:mdl}

The Bayesian formulation outlined above has an alternative --- but
entirely equivalent --- information-theoretic interpretation. We can
re-write the joint probability of Eq.~\ref{eq:joint} as
\begin{equation}
  P(\bm{A},\bm{k},\bm{e},\bm{b}) = 2^{-\Sigma}
\end{equation}
where
\begin{equation}
  \Sigma = -\log_2P(\bm{A},\bm{k},\bm{e},\bm{b}) = \mathcal{S} + \mathcal{L}
\end{equation}
is called the description length of the
data~\cite{rissanen_modeling_1978,grunwald_minimum_2007}, with
\begin{equation}
  \mathcal{S} = -\log_2P(\bm{A}|\bm{k},\bm{e},\bm{b})
\end{equation}
being the number of bits necessary to precisely describe the network, if
the model parameters are known, and
\begin{equation}
  \mathcal{L} = -\log_2P(\bm{k},\bm{e},\bm{b})
\end{equation}
being the number of bits necessary to describe the model
parameters. Hence, if we find the network partition that maximizes the
posterior of Eq.~\ref{eq:posterior}, we are automatically finding the
choice of parameters that \emph{most compresses} the data, i.e. yields
the shortest description length. This equivalence between Bayesian
inference and MDL holds much more
generally~\cite{grunwald_minimum_2007}, but with the microcanonical
formulation used here it is more directly evident.

The MDL interpretation also provides an intuitive explanation to why
this nonparametric approach is robust against overfitting: If the number
of groups becomes large, it will decrease $\mathcal{S}$ but increase
$\mathcal{L}$, with the latter functioning as a ``penalty'' that
disfavors overly complex models. For the same reason, the description
length can also be used as an application-independent criterion to
select between models of different classes, i.e. with a different
internal structure and set of parameters.  This type of comparison
amounts to a formal implementation of Occam's razor, where the simplest
model that can explain the data according to its statistical
significance should be selected (see also Sec.~\ref{sec:comparison}).

This equivalence means that other Bayesian approaches such as
Refs.~\cite{latouche_bayesian_2009,hofman_bayesian_2008,guimera_missing_2009,
yan_active_2010,come_model_2015,yan_bayesian_2016,newman_estimating_2016},
and those based on MDL,
e.g. Refs.~\cite{rosvall_information-theoretic_2007,peixoto_parsimonious_2013,
peixoto_hierarchical_2014}, correspond in fact to the same underlying
criterion. The main differences between those lie only in the actual
models used, the choice of priors, as well as more practical aspects
such as algorithmic complexity and approximations used.

However, it is important to emphasize that using either the Bayesian or
the MDL interpretation, we need to be open to the possibility that
different models --- or different parametrizations of the same model ---
may yield the same or very similar values for the description length or
posterior probability. In such situations, we should accept these
alternative explanations for the data on equal footing. The Bayesian
interpretation offers a more natural approach in these circumstances,
where instead of attempting to find the maximum of the posterior
distribution, we consider all possibilities, weighted according to their
posterior probability. This can be achieved by \emph{sampling} from the
posterior distribution using Monte Carlo techniques, as explained in
Sec.~\ref{sec:mcmc}.

When deciding which route to take --- to maximize or sample from the
posterior --- we need to acknowledge that therein lies the typical
trade-off between bias and variance: When maximizing the posterior, we
make a very specific statement about the data-generating process, but
which can include errors from many sources, such as lack of sufficient
statistics, degeneracy in the parameters or model misspecification. On
the other hand, when sampling from the posterior, we obtain results
which tend to be \emph{on average} less susceptible to those errors, but
which to the same degree are also more uncertain. Thus, we lose the
ability to make more specific assertions. Due to its nature, the latter
approach tends to incorporate more noise, and so the individual samples
run the risk of overfitting the data. Conversely, the maximization
approach tends to yield more conservative results, and thus runs the
risk of \emph{underfitting} the data, by omitting meaningful
features. Although in the ideal scenario where the model is well
specified and the data is plentiful both approaches must yield the same
result, in more realistic settings one source of error can only be
reduced at the expense of increasing the other. Hence, the final
decision must involve the ultimate objective of the inference task. In
general, we should expect sampling to be more suitable when the goal is
to generalize from the observed data and make predictions about new
measurements, wheres maximization tends to produce more accurate
representations of the observed data.

In Secs.~\ref{sec:comparison} and~\ref{sec:empirical} we compare results
obtained via strict MDL (i.e. maximization) and the Bayesian
(i.e. sampling) approaches on empirical data. In the following, we
proceed with defining the prior probabilities for the model
parameters. When discussing various possibilities, we will make use of
the MDL interpretation to decide which alternative yields the shortest
description for data that is more likely to be encountered.

\subsection{Prior for the node partition}

We begin with the prior for the partitions. Here we outline two general
approaches that will also be used for the remaining parameters. Firstly,
the simplest choice we could make is to be completely agnostic about
the partitions, and choose among all of them with equal probability,
\begin{equation}
  P(\bm{b}|B) = B^{-N}.
\end{equation}
However, this is not a good choice. The reason for this is that it
inherently assumes that the group sizes will be approximately the same,
since this is a typical property of completely random partitions.  Not
only is this unrealistic, but from a MDL perspective, whenever this is
not the case, we would miss an opportunity to further compress the
data. Therefore, we are better off instead replacing this by a
parametric distribution, that is conditioned on the group sizes
$\bm{n}=\{n_r\}$, where $n_r$ is the number of nodes in group $r$,
\begin{equation}\label{eq:micro_b}
  P(\bm{b}|\bm{n}) = \frac{\prod_rn_r!}{N!},
\end{equation}
which is a maximum entropy distribution (all allowed configurations are
equally likely), constrained on the fixed group sizes. In order to
remain nonparametric, we need a noninformative \emph{hyperprior} on the
node counts,
\begin{equation}\label{eq:hyper_n}
  P(\bm{n}|B) = \multiset{B}{N}^{-1},
\end{equation}
where $\multiset{n}{m}={n+m-1\choose m}$ counts the number of
$m$-combinations from a set of size $n$, or equivalently, the number of
possible histograms with $n$ bins with counts that sum to $m$. One may
argue, however, that the same principle should be applied again, with
the noninformative hyperprior above replaced by a parametric
distribution, with parameters sampled from a hyper-hyperprior, and so
on, indefinitely. However, proceeding like this yields increasingly
diminishing returns, and as we now show, there are good reasons to stop
at this point. If we take the logarithm of the joint probability
$P(\bm{b},\bm{n}|B)=P(\bm{b}|\bm{n})P(\bm{n}|B)$ and assume that the
groups are sufficiently large so that Stirling's factorial approximation
can be used, as well as $B\ll N$, we obtain
\begin{equation}
  \ln P(\bm{b},\bm{n}|B) \approx -NH(\bm{n}) - B\ln N
\end{equation}
where $H(\bm{n}) = -\sum_r(n_r/N)\ln(n_r/N)$ is the entropy of the group
size distribution. The first term in the equation above represents an
optimal limit, i.e. for sufficient data the negative log-probability
(the description length) approaches the entropy of the generating
distribution. Hence, if we were to replace the noninformative hyperprior
of Eq.~\ref{eq:hyper_n} with an even deeper Bayesian hierarchy, we would
gain at most a fairly marginal improvement proportional to $\ln N$,
which is unlikely to significantly alter the inference outcome.

The joint probability $P(\bm{b},\bm{n}|B)$ above has been used in
Refs.~\cite{latouche_bayesian_2009,peixoto_model_2015,come_model_2015,newman_estimating_2016},
but in some of these works it was equivalently derived as the marginal
distribution of the canonical model,
\begin{equation}\label{eq:canonical_b}
  P(\bm{b}|B) = \int P(\bm{b}|\bm{p})P(\bm{p}|B)\;\dd\bm{p}
\end{equation}
with
\begin{equation}
  P(\bm{b}|\bm{p}) = \prod_ip_{b_i}=\prod_rp_r^{n_r}
\end{equation}
where $p_r$ is the probability of a node belonging to group $r$, and
\begin{equation}
  P(\bm{p}|B) = (B-1)!\;\delta\left(1-\textstyle\sum_rp_r\right)
\end{equation}
is a uniform prior. Computing Eq.~\ref{eq:canonical_b} yields an
expression identical to
$P(\bm{b}|B)=P(\bm{b},\bm{n}|B)=P(\bm{b}|\bm{n})P(\bm{n}|B)$ using
Eqs.~\ref{eq:micro_b} and~\ref{eq:hyper_n} above.  However, there is an
apparently small detail that needs to be addressed. Namely, the maximum
entropy model of Eq.~\ref{eq:hyper_n} also generates groups with size
zero. This means that if we use it, we need to consider in our posterior
distributions partitions of the network that contain empty groups, which
would force us to treat the number of groups as a free variable that is
not necessarily equal to the number of observed (nonempty)
groups~\footnote{Note that this is not an issue when we are strictly
maximizing the posterior, since the most likely partition will never
contain empty groups.}. As shown in Ref.~\cite{newman_estimating_2016}
this requires a further complication of the inference algorithm, where
the number of groups is incorporated as a state variable. However, empty
groups possess no real value when interpreting the network structure:
saying that a network has five communities, but one of which is empty,
is the same as saying it has four communities, just in a more roundabout
and potentially misleading way. Hence, in order to avoid dealing with
such empty groups, and solving both of the above problems at once, we
simply exclude them from our prior distribution, by using instead
\begin{equation}\label{eq:hyper_nonempty}
  P(\bm{n}|B) = {N-1\choose B-1}^{-1}
\end{equation}
which is a uniform distribution over all histograms with $B$ bins and
counts that sum to $N$, where no bin is allowed to be empty.  With this
simple modification, the number of groups becomes a hard constraint as
well, and is always tied to the partition, thus obviating the need to
treat it as a free variable, and hence simplifying the inference
procedure. We note that while this modification is easy in the
microcanonical model, it is not as straightforward in the canonical
model of Eq.~\ref{eq:canonical_b}, since for every value of $p_r < 1$,
the probability that group $r$ will end up empty is strictly nonzero.

Lastly, we need a prior for the number of non-empty groups itself, which
we can choose as $P(B)=1/N$, for $B \in [1, N]$. (We could argue that,
since this amounts to a trivial multiplicative constant to the overall
probability, we could omit it completely. However, as it will be seen
further below, this term will not be a constant once we consider
hierarchical partitions.) With this, we have a nonparametric prior for
the partition that reads
\begin{align}
  P(\bm{b}) &= P(\bm{b}|\bm{n})P(\bm{n}|B)P(B) \nonumber\\
            &=\frac{\prod_rn_r!}{N!}{N-1\choose B-1}^{-1}\frac{1}{N}.
  \label{eq:partition_prior}
\end{align}
Since we are forbidding empty groups \emph{a priori},
from this point onward the value of $B$ will refer strictly to the
number of nonempty groups.

\subsection{Prior for the degrees}

\subsubsection{Non-degree-corrected model (NDC-SBM)}

We can recover a non-degree-corrected version of the microcanonical SBM
as a special case of the model above, by assuming that the half-edges
are randomly distributed among nodes of the same group, which yields a
particular probability for the degree sequence.

If at first we assume that all $e_r=\sum_se_{rs}$ half-edges incident on
group $r$ are distinguishable, they can be distributed among $n_r$ nodes
in $\Omega_r=n_r^{e_r}$ different ways. A particular degree sequence
inside group $r$ corresponds to exactly $\Xi_r(\bm{k})=e_r!/\prod_{i\in
r}k_i!$ such combinations, where the numerator accounts for the number
of permutations of half-edges, while the denominator discounts the
fraction of such permutations involving half-edges that are incident on
the same node, and hence amount to the same half-edge partition. The
probability of a particular degree sequence inside group $r$ is given by
the ratio $\Xi_r(\bm{k})/\Omega_r$, and thus the overall degree sequence
probability becomes
\begin{equation}
  P(\bm{k}|\bm{e},\bm{b}) = \prod_r\frac{e_r!}{n_r^{e_r}\prod_{i\in r}k_i!},
\end{equation}
which multiplied with Eq.~\ref{eq:model_dc} yields the model likelihood
\begin{equation}
  P(\bm{A}|\bm{e},\bm{b}) = \frac{\prod_{r<s}e_{rs}!\prod_re_{rr}!!}{\prod_rn_r^{e_r}}\times
  \frac{1}{\prod_{i<j}A_{ij}!\prod_iA_{ii}!!},
\end{equation}
which no longer depends explicitly on the degree sequence.

Like its canonical counterpart~\cite{karrer_stochastic_2011}, the
NDC-SBM will generate networks where nodes that belong to the same group
will have similar degrees, with a degree distribution inside each group
approaching asymptotically a Poisson. This means that the standard
deviation of the degrees inside group $r$ will be $\sigma_k =
\sqrt{\avg{k}_r}$, with $\avg{k}_r=e_r/n_r$ being the average degree. As
argued in Ref.~\cite{karrer_stochastic_2011}, this is an unrealistic
assumption for many empirical networks, most of which possess very
heterogeneous degree distributions. As a result, attempts to infer the
SBM on such networks can amount largely to a division of the nodes into
degree classes. It is therefore useful to postulate prior probabilities
that can account for arbitrary degree sequences, as we do in the
following.

\subsubsection{Arbitrary degree sequences}

Similarly to the partition of the nodes, the simplest choice we can make
is to sample the degrees inside each group from a uniform distribution,
\begin{equation}\label{eq:k_uniform_prior}
  P(\bm{k}|\bm{e},\bm{b}) = \prod_r\multiset{n_r}{e_r}^{-1}
\end{equation}
where $\multiset{n_r}{e_r}$ counts the number of possible degree
sequences on $n_r$ nodes, constrained such that their total sum equals
$e_r$. But again, such a uniform assumption is not the best choice: If
we sample from this prior, we still obtain degree sequences where most
nodes have very similar degrees. Indeed, if the number of nodes is
sufficiently large, it can be shown that the expected degree
distribution inside each group with the above prior will approach an
exponential $p_k = p(1-p)^k$, with an average $\avg{k}=(1-p)/p$ (see
Appendix~\ref{sec:deg_dist}). The expected standard deviation is
therefore $\sigma_k = \sqrt{1-p}/p = O(\avg{k})$, which, although larger
than what is obtained with the NDC-SBM, is still significantly smaller
that expected for many empirical networks~\cite{clauset_power-law_2009}.

\begin{figure}
  \includegraphics[width=\columnwidth]{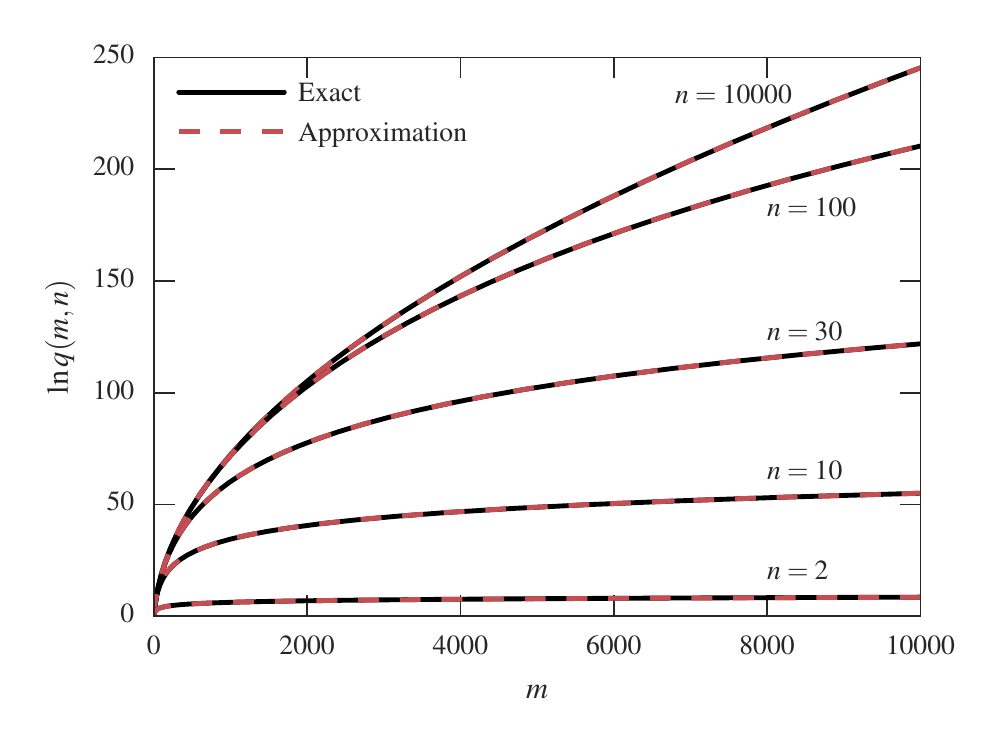}\\
  \includegraphics[width=\columnwidth]{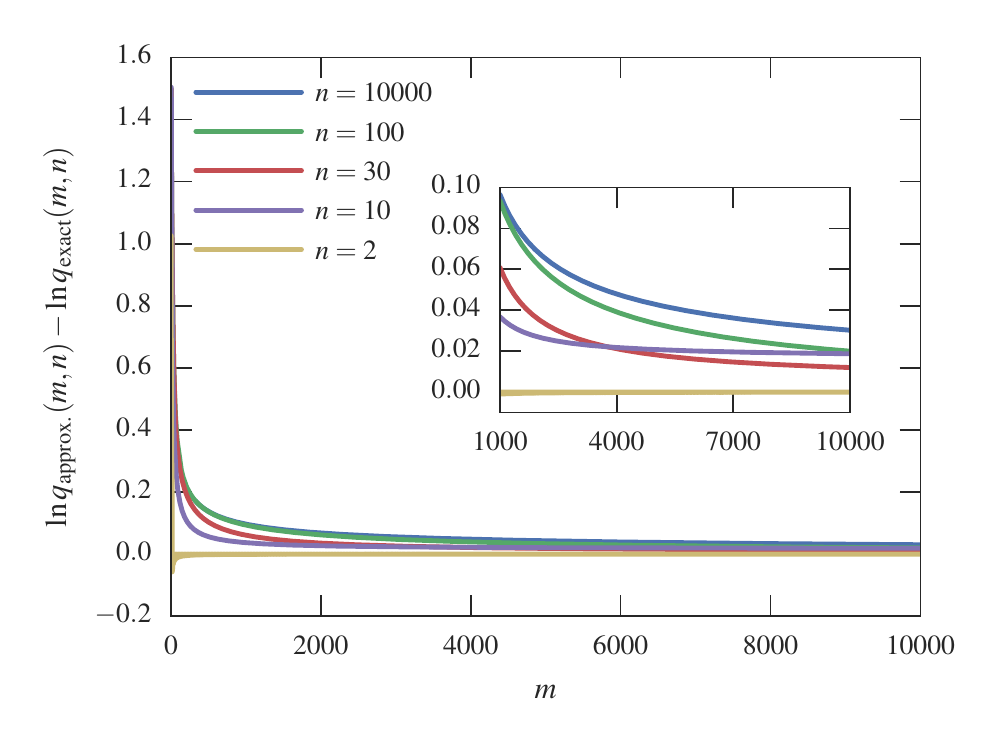}

  \caption{Comparisons between the exact and approximated values of the
  number of restricted partitions $q(m,n)$, using Eqs.~\ref{eq:q-exact}
  and \ref{eq:q-approx} to~\ref{eq:erdos-lehner}. The top panel shows
  both values computed for different values of $m$ and $n$, and the
  bottom panel shows the absolute difference of their logarithms, with
  the inset displaying a zoom into the large $m$ region.
  \label{fig:q-approx}}
\end{figure}

In view of this, and following the same logic employed for the node
partition, a better prior for $\bm{k}$ should be conditioned on an
arbitrary degree distribution $\bm{\eta}=\{\eta_k^r\}$, with $\eta_k^r$
being the number of nodes with degree $k$ that belong to group $r$,
\begin{equation}\label{eq:k_dist}
  P(\bm{k}|\bm{e},\bm{b}) = P(\bm{k}|\bm{\eta})P(\bm{\eta}|\bm{e},\bm{b})
\end{equation}
and where
\begin{equation}
  P(\bm{k}|\bm{\eta}) = \prod_r \frac{\prod_k\eta_k^r!}{n_r!}
\end{equation}
is a uniform distribution of degree sequences constrained by the overall
degree counts, and
\begin{equation}\label{eq:k_dist_prior}
  P(\bm{\eta}|\bm{e},\bm{b}) = \prod_r q(e_r, n_r)^{-1}
\end{equation}
is the distribution of the overall degree counts. The quantity $q(m,n)$ is
the number of different degree counts with the sum of degrees being
exactly $m$ and that have at most $n$ non-zero counts. This is also
known as the number of \emph{restricted partitions} of the integer $m$
into at most $n$ parts~\cite{andrews_theory_1984}. The function $q(m,
n)$ can be computed exactly via the recurrence
\begin{equation}\label{eq:q-exact}
  q(m, n) = q(m, n - 1) + q(m - n, n),
\end{equation}
and the boundary conditions $q(m, 1) = 1$ for $m>0$, and $q(m, n) = 0$
for $m\le 0$ or $n\le 0$.  With this, the full table of values for $m
\le M$ and $n \le m$ can be computed in time $O(M^2)$. Hence, if the
number of edges and nodes is not too large, we can pre-compute these
values as a setup to the inference procedure. However, this can still
become computationally expensive for very large systems. Unfortunately,
no closed-form expression for $q(m, n)$ is known which would allow us to
compute it in constant time. Fortunately, however, accurate asymptotic
expressions are known, which permit efficient computation for large
arguments. Namely, for large values of $m$ the number of partitions
approaches asymptotically the following
value~\cite{szekeres_asymptotic_1951,szekeres_asymptotic_1953,canfield_recursions_1996}
\begin{equation}\label{eq:q-approx}
  q(m, n) \approx \frac{f(u)}{m}\exp(\sqrt{m}g(u)),
\end{equation}
where $u=n/\sqrt{m}$ and the functions $f(u)$ and $g(u)$ are given by
\begin{align}
  f(u) &= \frac{v(u)}{2^{3/2}\pi u}\left[1-(1+u^2/2)e^{-v(u)}\right]^{-1/2},\\
  g(u) &= \frac{2v(u)}{u} - u\ln(1-e^{-v(u)}),
\end{align}
and $v(u)$ is given implicitly by solving
\begin{equation}\label{eq:v}
  v = u\sqrt{-v^2/2 - \operatorname{Li}_2(1-e^v)},
\end{equation}
where $\operatorname{Li}_2(z) = -\int_0^z[\ln(1-t)/t]\dd t$ is the
dilogarithm function. (Eq.~\ref{eq:v} can be easily solved numerically
via Newton's method, or simply via repeated iteration, which converges
within machine precision usually after only very few steps). This
approximation holds for values of $n \ge m^{1/6}$. For smaller values $n
\ll m^{1/3}$ we have instead~\cite{erdos_distribution_1941}
\begin{equation}\label{eq:erdos-lehner}
  q(m, n) \approx \frac{{m - 1 \choose n - 1}}{m!}.
\end{equation}
With Eqs.~\ref{eq:q-approx} to~\ref{eq:erdos-lehner} we have an
approximation for $q(m,n)$ for the entire range of parameters $m$ and
$n$ that is remarkably accurate, as shown in Fig.~\ref{fig:q-approx}:
For arguments of the order $10^3$, the largest log ratio between the
approximate and exact values is only around $0.1$, which has a
negligible effect on the outcome of hypothesis testing, and is below the
accuracy usually required for MCMC sampling.  In our implementation, we
pre-compute $q(m,n)$ using the exact Eq.~\ref{eq:q-exact} for $m <
10^4$, and resort to Eqs.~\ref{eq:q-approx} to~\ref{eq:erdos-lehner}
only for larger arguments, thus guaranteeing a computation of $q(m,n)$
in time $O(1)$, and hence incurring a negligible impact in the overall
algorithmic complexity of the inference procedure.

\begin{figure}
  \includegraphics[width=\columnwidth]{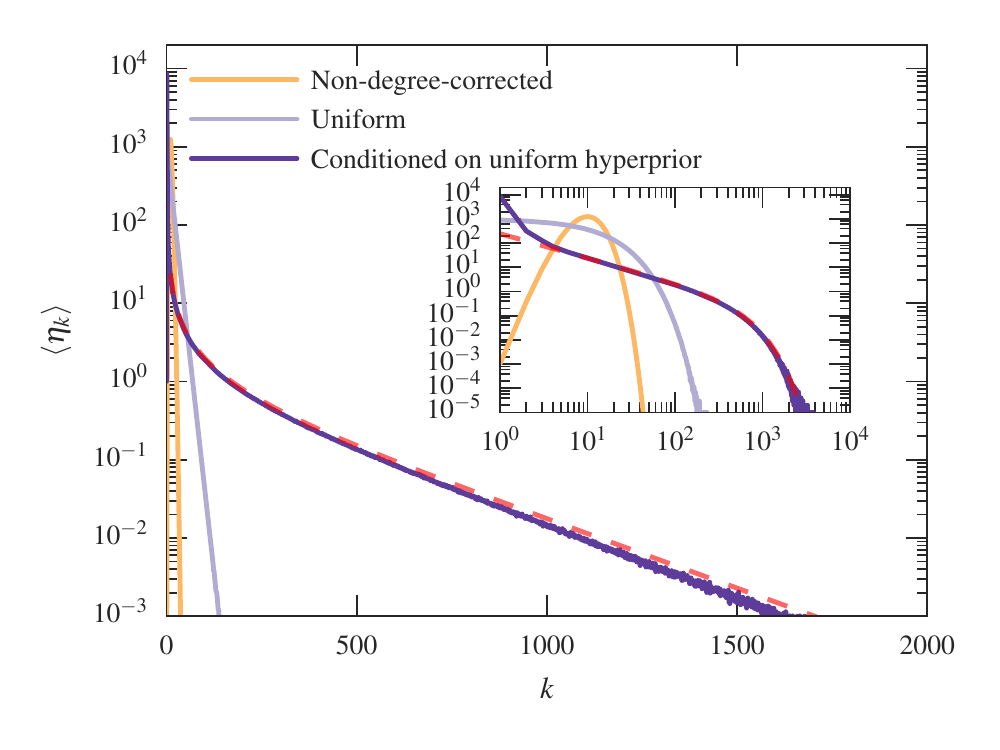}

  \caption{Expected degree distributions for the three different priors
  considered in the text for the degree sequence inside each group ---
  the NDC-SBM, the uniform prior of Eq.~\ref{eq:k_uniform_prior} and the
  prior of Eq.~\ref{eq:k_dist_prior} conditioned on a degree
  distribution sampled randomly --- for $N=10^4$ nodes and average
  degree $\avg{k}=10$. In all cases, the distributions were sampled from
  their respective microcanonical distributions using rejection
  sampling. The dashed line shows the Bose-Einstein distribution of
  Eq.~\ref{eq:bose-einstein}. \label{fig:avg_hist}}
\end{figure}

As seen in Fig.~\ref{fig:avg_hist}, the expected degree distribution
sampled from Eq.~\ref{eq:k_dist_prior} is typically significantly
broader than the exponential distribution obtained with
Eq.~\ref{eq:k_uniform_prior}. As shown in Appendix~\ref{sec:deg_dist},
this will approach a Bose-Einstein distribution, with a variance
$\sigma_k^2 \propto \sqrt{N}$ that will diverge for a large system
size. In particular, the distribution will asymptotically approach a
scale-free form $p_k \sim 1/k$ for $k \ll \sqrt{E}$, followed by an
exponential decay for larger arguments.

Although this prior assumption clearly favors broader degree
distributions, it could be argued that it still does not properly
capture the structure of real networks, most of which also do not
possess a Bose-Einstein degree distribution. Indeed, it may seem that by
changing between the priors considered above, we have simply switched
between Poisson, geometric and Bose-Einstein distributions, which are
just three of an infinite range of possibilities. However, in reality,
the conditioned prior of Eq.~\ref{eq:k_dist} will not concentrate as
strongly on the expected distribution as the other two, and thus will
not significantly penalize distributions that deviate from it, even if
the deviation is very large, as will now be shown.

In order to assess the improvement brought on by the conditioned prior,
it is instructive to obtain the asymptotic behavior of $q(m,n)$ in the
limit of ``sufficient data'' with $m\gg 1$ and $n \gg 1$, which is given
by~\cite{erdos_distribution_1941}
\begin{equation}
  q(m, n) \approx p(m) \exp\left(-\frac{\sqrt{6m}}{\pi}e^{-\pi n/\sqrt{6m}}\right),
\end{equation}
as long as $n \gg \sqrt{m}$ and where $p(m)=q(m,m)$ is the number of
unconstrained partitions of $m$, which itself is given exactly by the
recursion
\begin{equation}
  p(m)=\sum_{k > 0} (-1)^{k-1}p\left(m- k(3k -1)/2\right)
\end{equation}
and for large values of $m$ by the Hardy-Ramanujan
formula~\cite{hardy_formule_1917,hardy_asymptotic_1918}
\begin{equation}\label{eq:hardy-ramanujan}
  p(m) \approx \frac{1}{4\sqrt{3}m} \exp(\pi\sqrt{2m/3}).
\end{equation}
With these results, we see immediately that for ``sparse'' groups with
$e_r \propto n_r$ and $n_r\gg 1$ we have $\ln q(e_r, n_r) \sim
O(\sqrt{n_r})$, and hence
\begin{equation}\label{eq:k_dist_asym}
  \ln P(\bm{k}|\bm{e},\bm{b}) \approx -\sum_rn_rH(\bm{\eta}_r) + O(\sqrt{n_r}),
\end{equation}
where $H(\bm{\eta}_r) = -\sum_k(\eta_k^r/n_r)\ln(\eta_k^r/n_r)$ is the
entropy of the empirical degree distribution in group $r$. Therefore,
for sufficiently many nodes in each group, the hyperprior of
Eq.~\ref{eq:k_dist_prior} will ``wash out'' and the probability of
Eq.~\ref{eq:k_dist} will approach that of the the \emph{actual} degree
sequence, whatever its form may be, even if it deviates from the typical
form of Fig.~\ref{fig:avg_hist}. This is not the case of the uniform
prior of Eq.~\ref{eq:k_uniform_prior}, which is not able to ``learn''
the underlying distribution in the same manner. Eq.~\ref{eq:k_dist_asym}
also means that an exact prior knowledge of the \emph{true} degree
distribution in each group would improve the log-probability (and the
description length) only by a factor $O(\sqrt{n_r})$, which will be
dwarfed asymptotically by the remaining terms that scale linearly as
$O(n_r)$. Therefore, any further improvement in the choice of prior for
the degree sequence is confined to a relatively narrow range, similarly
to what happens with the prior for the partition of the nodes into
groups.

\subsection{Prior for the edge counts and nested SBM hierarchies}

The remaining piece is the prior for the edge counts between groups,
$\bm{e}$.  We can start again with a uniform prior
\begin{equation}\label{eq:uniform_edge_prior}
  P(\bm{e}) = \multiset{\multiset{B}{2}}{E}^{-1},
\end{equation}
where $\multiset{\multiset{B}{2}}{E}$ counts the number of symmetric
$e_{rs}$ matrices with a constrained sum $\sum_{rs}e_{rs}=2E$.

Perhaps unsurprisingly at this point, this is also not a good
choice. This time, however, the negative effects are somewhat more
dramatic than the previous choices of uniform priors. Namely, this
assumption will limit our capacity to detect small groups in very large
networks: It introduces a ``resolution limit'', where the largest number
of groups that can be inferred scales as
$B_{\text{max}}\sim\sqrt{N}$~\cite{peixoto_parsimonious_2013}, similar
to what is observed with the modularity maximization
heuristic~\cite{fortunato_resolution_2007}. We revisit this issue in
more detail in Sec.~\ref{sec:resolution}.

As was shown in Ref.~\cite{peixoto_hierarchical_2014}, this problem can
be solved again by deepening the Bayesian hierarchy. It is useful now to
notice that the matrix $\bm{e}$ can be interpreted as the adjacency
matrix of a multigraph with $B$ nodes and $E$ edges. Hence, an
appropriate choice seems to be to use the SBM again to generate it,
where each group $r$ belongs to one of another set of groups, and so on
recursively, a $L$ number of times,
\begin{equation}\label{eq:upper}
  P(\{\bm{e}_l\}|\{\bm{b}_l\}) = \prod_{l=1}^LP(\bm{e}_l|\bm{e}_{l+1},\bm{b}_l),
\end{equation}
where $\bm{b}_l$ is the partition of the groups in level $l$, $\bm{e}_l$
is the (weighted) adjacency matrix at level $l$, and we enforce always
that $B_L=1$. Note that since the number of edges is the same in all
levels while the number of nodes decreases, the multigraphs become
increasingly denser at the upper levels, and the occurrence of parallel
edges becomes predominant, even if the graph at the lowest level is
sparse and simple. Although the likelihood of Eq.~\ref{eq:model_dc} that
was used at the bottom level also admits arbitrarily dense multigraphs,
it will not generate them uniformly within the SBM constraints, since it
is based on an uniform generation of \emph{configurations}. Because of
this, it is not a good idea to use the exact same model as the priors in
the upper layers, which will introduce a significant bias as the
multigraphs become dense. Indeed, simply inserting Eq.~\ref{eq:model_dc}
into Eq.~\ref{eq:upper} makes all successive levels cancel out in the
likelihood, yielding a trivial model where only the first and last
levels have any contribution. A much better approach, which is unbiased
and maximally noninformative within the imposed constraints, is to use a
uniform NDC-SBM for multigraphs directly, where all allowed multigraphs
(not their corresponding configurations) occur with the same
probability. The likelihood can be obtained via basic
enumeration~\cite{peixoto_entropy_2012}, and is given by
\begin{equation}\label{eq:multi_sbm}
  P(\bm{e}_l|\bm{e}_{l+1},\bm{b}_l) = \prod_{r<s}\multiset{n_r^ln^l_s}{e_{rs}^{l+1}}^{-1}
  \prod_{r}\multiset{n_r^l(n_r^l+1)/2}{e_{rr}^{l+1}/2}^{-1}.
\end{equation}
Note that if we make $L=1$, we recover the uniform prior of
Eq.~\ref{eq:uniform_edge_prior}, making it a special case. To complete
the model, we need also the prior for the partitions in all levels,
\begin{equation}\label{eq:prior_hierarchy}
P(\{\bm{b}_l\}) = \prod_{l=1}^LP(\bm{b}_l),
\end{equation}
where for each level we use again Eq.~\ref{eq:partition_prior}, but
replacing $B\to B_l$ and $N\to B_{l-1}$, with the boundary condition
$B_0=N$.

The depth $L$ of the hierarchy itself is something that we want to infer
from the data as well. One approach, for instance, is to put a
noninformative prior on it $P(L)=1/L_{\text{max}}$, with some maximum
possible value $L_{\text{max}}$ that is sufficiently large,
e.g. $L_{\text{max}}=N$. But since this contributes to nothing but an
overall multiplicative constant in the distribution, it can be omitted
altogether.

\subsection{Model summary}

Putting together the model likelihood with all the priors, we have a
joint distribution for the hierarchical microcanonical DC-SBM that reads
\begin{widetext}
\begin{align}
  P(\bm{A},\bm{k},\{\bm{e}_l\},\{\bm{b}_l\}) &= P(\bm{A}|\bm{k},\bm{e},\bm{b}_1)\times P(\bm{k}|\bm{e}_1,\bm{b}_1)\times P(\{\bm{e}_l\})\times P(\{\bm{b}_l\})\label{eq:full_joint}\\
  &= \frac{\prod_ik_i!\prod_{r<s}e_{rs}!\prod_re_{rr}!!}{\prod_re_r!\prod_{i<j}A_{ij}!\prod_iA_{ii}!!}\times
  \prod_r\frac{\prod_k\eta_k^r!}{n_r!} q(e_r, n_r)^{-1}\times \nonumber\\
  &\qquad\prod_{l=1}^L\prod_{r<s}\multiset{n_r^ln^l_s}{e_{rs}^{l+1}}^{-1}
  \prod_{r}\multiset{n_r^l(n_r^l+1)/2}{e_{rr}^{l+1}/2}^{-1}\times
  \frac{\prod_rn_r^l!}{B_{l-1}!}{B_{l-1}-1\choose B_l-1}^{-1}\frac{1}{B_{l-1}}.\label{eq:full_joint_big}
\end{align}
\end{widetext}
It is important to emphasize that this likelihood has the following
useful property: When considering the difference in the log-likelihood
after moving a single node $i$ from a group to another, it is necessary
only to consider a number of terms that is proportional to the number of
groups that are involved in the change, i.e. those of the node that is
being moved and its neighbors. Therefore, in the worse case, we need to
update $O(k_i)$ terms, a number which is \emph{independent} of the total
number of groups in the bottom of the hierarchy, $B_1$. This contrasts
with other formulations that require the computation of a number of
terms that is linearly proportional to the total number of groups
(e.g.~\cite{guimera_missing_2009,yan_active_2010,come_model_2015,newman_estimating_2016}),
or even quadratic (e.g.~\cite{decelle_inference_2011}). This property
will permit the inference on large networks, for which the appropriate
number of groups is likely to be large as well, as we describe in
Sec.~\ref{sec:mcmc}.

In addition to this model, the NDC-SBM and the alternative version of
the DC-SBM with uniform priors on the degrees can be obtained simply by
replacing the prior $P(\bm{k}|\bm{e},\bm{b}_0)$ in
Eq.~\ref{eq:full_joint} with the appropriate one. This does not change
the efficiency of the likelihood computation described
above. Furthermore, as mentioned previously, the non-hierarchical
version of each model can be recovered by simply enforcing a hierarchy
with just one level, i.e. $L=1$.

\section{Ensemble equivalence}\label{sec:canonical}

The microcanonical model above differs from the most common
``canonical'' formulation of the SBM, where the modular network
structure is imposed via ``soft'' constraints, that are obeyed only on
average. For example, the original canonical Poisson formulation of the
DC-SBM~\cite{karrer_stochastic_2011} is
\begin{widetext}
\begin{align}\label{eq:canonical_dcsbm}
  P(\bm{A}|\bm{\lambda},\bm{\theta}) &= \prod_{i<j} \frac{(\theta_i\theta_j\lambda_{b_ib_j})^{A_{ij}}e^{-\theta_i\theta_j\lambda_{b_ib_j}}}{A_{ij}!}\prod_i \frac{(\theta_i^2\lambda_{b_ib_i}/2)^{A_{ij}/2}e^{-\theta_i^2\lambda_{b_ib_i}/2}}{(A_{ii}/2)!}\nonumber\\
                                   &= \prod_{r<s}\lambda_{rs}^{e_{rs}}e^{-\lambda_{rs}\hat\theta_r\hat\theta_s} \prod_r\lambda_{rr}^{e_{rr}/2}e^{-\lambda_{rr}\hat\theta_r^2/2}\times\frac{\prod_i\theta_i^{k_i}}{\prod_{i<j}A_{ij}!\prod_iA_{ii}!!}.
\end{align}
\end{widetext}
where $\theta_i$ determines the propensity of node $i$ to receive edges,
whereas $\lambda_{rs}$ controls the distribution of edges between
groups, and with
\begin{equation}
  \hat\theta_r = \sum_i\theta_i\delta_{b_i,r}.
\end{equation}
In this model, the degrees of the nodes and the number of edges between
groups are fixed only in expectation, but otherwise can fluctuate
between samples. If one applies Stirling's factorial approximation $\ln
m! \approx m\ln m - m$ to the terms of Eqs.~\ref{eq:omega}
and~\ref{eq:xi} that depend on $e_{rs}$ and $k_i$, it is easily seen
that the microcanonical likelihood of Eq.~\ref{eq:model_dc} approaches
Eq.~\ref{eq:canonical_dcsbm}, which means both models generate the same
networks with the same probability asymptotically, if the parameters are
chosen in a compatible manner, e.g. $\theta_i = k_i/e_{b_i}$ and
$\lambda_{rs}=e_{rs}$. However, this only holds if the edge counts
between groups \emph{as well as} the degrees of the nodes become
sufficiently large. For smaller or sparser networks, on the other hand,
the differences can be important, and it is well understood that the
microcanonical and canonical ensembles are not equivalent in these
cases~\cite{bianconi_entropy_2009,peixoto_entropy_2012,squartini_breaking_2015,garlaschelli_ensemble_2016}.
However, an exact equivalence between these ensembles can in fact be
obtained in a Bayesian setting, via the computation of the marginal
likelihood that involves integrating over the canonical parameters,
$\bm{\theta}$ and $\bm{\lambda}$, weighted with a prior probability, as
will now be shown.

Before we can proceed with the computation of the marginal likelihood,
we must notice that the model parameters are determined only up to an
arbitrary multiplicative constant, since the likelihood of
Eq.~\ref{eq:canonical_dcsbm} depends only on their products
$\theta_i\theta_j\lambda_{b_ib_j}$. Although their absolute values are
in principle arbitrary, the exact parametrization we choose will affect
the choice of priors we can make, and ultimately the marginal
likelihood. Here we will contrast two possible choices. We begin with
the assumption made in Refs.~\cite{yan_bayesian_2016,newman_estimating_2016}
\begin{equation}\label{eq:theta_v1}
  \hat\theta_r = n_r.
\end{equation}
If we make this choice, the value of $\lambda_{rs}$ corresponds to the
average probability of two nodes in groups $r$ and $s$ being
connected. We can then choose a noninformative prior for $\bm{\lambda}$,
conditioned only on the expected density of the network, $p=2E/N^2$,
\begin{equation}\label{eq:lambda_prior_1}
  P(\lambda_{rs}) = e^{-\lambda_{rs}/p}/p.
\end{equation}
For $\bm{\theta}$, we use also a noninformative distribution,
\begin{equation}
  P(\bm{\theta}|\bm{b}) = \prod_r\frac{(n_r-1)!}{n_r^{n_r-1}}\;\delta(\hat\theta_r-n_r),
\end{equation}
subject only to the scaled simplex constraint of
Eq.~\ref{eq:theta_v1}. As computed in Ref.~\cite{newman_estimating_2016},
the marginal likelihood is therefore,
\begin{widetext}
\begin{align}
  P_1(\bm{A}|\bm{b}) &= \int P(\bm{A}|\bm{\lambda},\bm{\theta}) P(\bm{\lambda}) P(\bm{\theta}|\bm{b})\,\dd\bm\lambda\,\dd\bm\theta\nonumber\\
  &= p^E\prod_{r<s}\frac{e_{rs}!}{(pn_rn_s+1)^{e_{rs}+1}}\prod_r\frac{(e_{rr}/2)!}{(pn_r^2/2+1)^{e_{rs}/2+1}}\prod_r\frac{n_r^{e_r}(n_r-1)!}{(e_r+n_r-1)!}\frac{\prod_ik_i!}{\prod_{i<j}A_{ij}!\prod_iA_{ii}!!}.\label{eq:canonical}
\end{align}
\end{widetext}
This marginal likelihood is not equivalent to the microcanonical model
presented previously, and hence corresponds to a different overall
generative process. However, things are different if we assume another
parametrization, namely
\begin{equation}\label{eq:theta_v2}
  \hat\theta_r = 1.
\end{equation}
In this case, the value of $\lambda_{rs}$ represents the average number
of edges between groups $r$ and $s$ (or twice that for $r=s$). Similar
to the previous case, we can choose a noninformative prior for
$\bm{\lambda}$, conditioned only on the expected total number of edges,
\begin{equation}\label{eq:lambda_prior_2}
  P(\lambda_{rs}) =
  \begin{cases}
    e^{-\lambda_{rs}/\bar\lambda}/\bar\lambda &\text{ if } r\neq s,\\
    e^{-\lambda_{rs}/2\bar\lambda}/2\bar\lambda &\text{ if } r = s,
  \end{cases}
\end{equation}
with $\bar\lambda = 2E/B(B+1)$. Like before, for $\bm{\theta}$ we
use noninformative distribution,
\begin{equation}
  P(\bm{\theta}|\bm{b}) = \prod_r(n_r-1)!\;\delta(\hat\theta_r-1),
\end{equation}
but subject now to the simplex constraint of Eq.~\ref{eq:theta_v2}
instead. Performing the same integral, the marginal likelihood then
becomes
\begin{widetext}
  \begin{equation}\label{eq:canonical_alt}
    P_2(\bm{A}|\bm{b}) = \frac{\bar\lambda^E}{(\bar\lambda+1)^{E+B(B+1)/2}}\times\frac{\prod_{r<s}e_{rs}!\prod_r e_{rr}!!}{\prod_{i<j}A_{ij}!\prod_i A_{ii}!!} \prod_r\frac{(n_r-1)!}{(e_r+n_r-1)!}\prod_ik_i!,
  \end{equation}
\end{widetext}
from which we can immediately recognize the microcanonical model by
re-writing the likelihood as
\begin{equation}
  P_2(\bm{A}|\bm{b}) = P(\bm{A}|\bm{k},\bm{e},\bm{b})P(\bm{k}|\bm{e},\bm{b})P(\bm{e}),
\end{equation}
where $P(\bm{A}|\bm{k},\bm{e},\bm{b})$ is the microcanonical likelihood
of Eq.~\ref{eq:model_dc}, $P(\bm{k}|\bm{e},\bm{b})$ is the
noninformative degree-sequence probability of
Eq.~\ref{eq:k_uniform_prior} and $P(\bm{e})$ is the probability of the
degree counts as $B(B+1)/2$ independent exponential variables with
average $\bar\lambda$,
\begin{align}
  P(\bm{e}) &= \prod_{r<s}(1-\mu)^{e_{rs}}\mu  \prod_{r}(1-\mu)^{e_{rr}/2}\mu \\
  &=\bar\lambda^E/(\bar\lambda+1)^{E+B(B+1)/2},
\end{align}
where $\mu = 1 / (\bar\lambda + 1)$. This last prior $P(\bm{e})$ is
different from the microcanonical one used in
Eq.~\ref{eq:uniform_edge_prior} simply in that here the total number of
edges is allowed to fluctuate, being constrained only in
expectation. Otherwise, the likelihoods of the canonical and
microcanonical models are identical. This means that although both
formulations involve distinct generative processes, these are not in
fact distinguishable from data. This is fortunate, since it eliminates
at least one arbitrary choice we have to make prior to inferring the
modular structure of networks, and shows that the choice of ensemble can
be largely subjective.

However, we are still left with a seemingly arbitrary choice of
parametrization, having to decide between Eq.~\ref{eq:theta_v1} (option
1) and Eq.~\ref{eq:theta_v2} (option 2). As the results above show,
these choices correspond to different assumptions about the
data-generating process. In the first case, the expected number of edges
between groups $r$ and $s$ (according to the prior for $\bm{\lambda}$)
is assumed to depend on the sizes of the groups,
i.e. $\avg{e_{rs}}=n_rn_sp$. This is the same expected value for the
same partition of a completely random network with density $p$. In the
second case, however, this value is independent of the group sizes
$\avg{e_{rs}}=\bar{\lambda}$, and deviates from the expected fully
random value whenever the groups sizes are not the same. Hence, the
ensembles generated in each case are indeed different, and to decide
which one should be used is a model selection problem. As will be
discussed in more detail in Sec.~\ref{sec:comparison}, this can be
performed by inspecting the marginal likelihood ratio between both
models, assuming the same node partition,
\begin{equation}
  \Lambda = \frac{P_2(\bm{A}|\bm{b})}{P_1(\bm{A}|\bm{b})}
\end{equation}
where $P_1(\bm{A}|\bm{b})$ and $P_2(\bm{A}|\bm{b})$ correspond to
Eqs.~\ref{eq:canonical} and~\ref{eq:canonical_alt}, respectively. If
we assume $N \gg B^2$, this ratio amounts to a simple expression
\begin{equation}
  \ln\Lambda \approx \sum_{r\ge s} \left[\frac{e_{rs}}{pn_rn_s} - \ln \frac{(1+\delta_{rs})N^2}{B(B+1)n_rn_s} - 1\right].
\end{equation}
From this, and if we further assume groups of equal sizes $n_r=N/B$ as
well as $B\gg 1$, we see that as the network approaches a fully random
structure with $e_{rs} = pn_rn_s$, we have $\ln\Lambda \to -B\ln 2$ and
hence a situation that favors option 1. However, as the data become more
structured, this is more often not the case. This is better seen by
considering a special case known as the planted partition
model~\cite{condon_algorithms_2001}, composed of $B$ equal-sized groups
and edge counts given by
\begin{equation}
  e_{rs}=2E\left[\frac{c}{B}\delta_{rs} + \frac{(1-c)}{B(B-1)}(1-\delta_{rs})\right],
\end{equation}
with $c \in [0, 1]$ controlling the degree of assortativity.
Substituting this in the above, we have
\begin{equation}
  \ln\Lambda \approx \frac{B^2(c + 1)}{2} - \frac{B(B+1)}{2}\ln\left(\frac{eB}{B+1}\right) - B\ln 2,
\end{equation}
which is independent of the size of the network, and grows only with the
number of groups and assortativity. For $B \gg 1$, we have $\ln\Lambda >
0$ if $c > (2\ln 2) / B \approx 1.4/B$. The ensemble is equivalent to a
fully random network at a slightly smaller value $c = 1/B$ [but is
already undetectable at $c=1/B \pm (B -
1)/(B\sqrt{\avg{k}})$~\cite{decelle_asymptotic_2011}]. Hence, as the
number of groups increases, for the vast majority of parameter choices
$c\in [(2\ln 2) / B, 1]$ we have that option 2 is favored with a
confidence that grows as $\ln\Lambda = O(B^2)$.

Beside these arguments, there are other more important reasons to prefer
option 2. If we adopt its microcanonical interpretation, we can address
the issues with the noninformative priors discussed in the previous
sections, and replace both $P(\bm{k}|\bm{e},\bm{b})$ and $P(\bm{e})$ by
distributions conditioned on hyperparameters. Furthermore, as already
mentioned, changes to the likelihood of Eq.~\ref{eq:canonical_alt} can be
computed more efficiently than Eq.~\ref{eq:canonical}: If we move a node
$i$ to a new group, we need to update $O(B)$ terms in
Eq.~\ref{eq:canonical}, whereas in Eq.~\ref{eq:canonical_alt} at most
only $O(k_i)$ terms need to be recomputed (independent of $B$). This
leads to a substantial improvement in the performance of inference
algorithms, as discussed further in Sec.~\ref{sec:mcmc}.

\section{How many groups can be inferred?}\label{sec:resolution}

One of the main strengths of the nonparametric approach presented here
is that it can be used to determine the number of groups $B$, in
addition to the other model parameters. One natural question that arises
is whether there are intrinsic limitations associated with the inference
of this parameter. In particular, here we are interested in the
situation where the inferred number of groups $B^*$ is smaller than the
true value $B$ used to generated the network, such that parts of the
modular structure are not resolved by inference. As shown in
Ref.~\cite{peixoto_parsimonious_2013} with a simplified version of the
model presented here, if the size and density of the network are kept
fixed, and the planted value exceeds a threshold $B>B_{\text{max}}$, we have that
$B^*=B_{\text{max}}$ and the planted modular structure cannot be fully resolved. In
particular, the choice of a noninformative prior for the edge counts
$P(\bm{e})$ leads to a limitation where at most only $B_{\text{max}} = O(\sqrt{N})$
groups can be identified. Replacing this noninformative prior by a
series of nested SBMs was shown in Ref.~\cite{peixoto_hierarchical_2014}
to significantly alleviate this limitation, increasing the maximum
number of groups to $B_{\text{max}} = O(N/\ln N)$. Here we revisit this issue,
considering the more elaborate models presented in this work.

We perform our analysis on a degree-corrected planted partition model,
with $B$ planted groups of equal size, each containing exactly $E/B$
edges connecting their nodes randomly, and no connections at all between
nodes of different groups, i.e. $e_{rs}=2E\delta_{rs}/B$. The likelihood
of any particular network sampled from this model is
\begin{equation}\label{eq:pp_l}
  P(\bm{A}|\bm{k},\bm{e},\bm{b}) = \frac{(2E/B)!!^{B}}{(2E/B)!^B}\times\frac{\prod_ik_i!}{\prod_{i<j}A_{ij}!\prod_iA_{ii}!!},
\end{equation}
and with prior probabilities
\begin{align}
  P(\bm{b}) &= \frac{(N/B)!^{B}}{N!}\times {N-1\choose B-1}^{-1}\frac{1}{N},\label{eq:pp_b}\\
  P(\bm{e}|\bm{b}) &= \multiset{B(B+1)/2}{E},\label{eq:pp_e}\\
  P(\bm{k}|\bm{e},\bm{b}) &= \multiset{N/B}{2E/B}^{-B}\label{eq:pp_k},
\end{align}
where we have used the noninformative priors for the edge counts and
degrees.

\begin{figure}
  \includegraphics[width=\columnwidth]{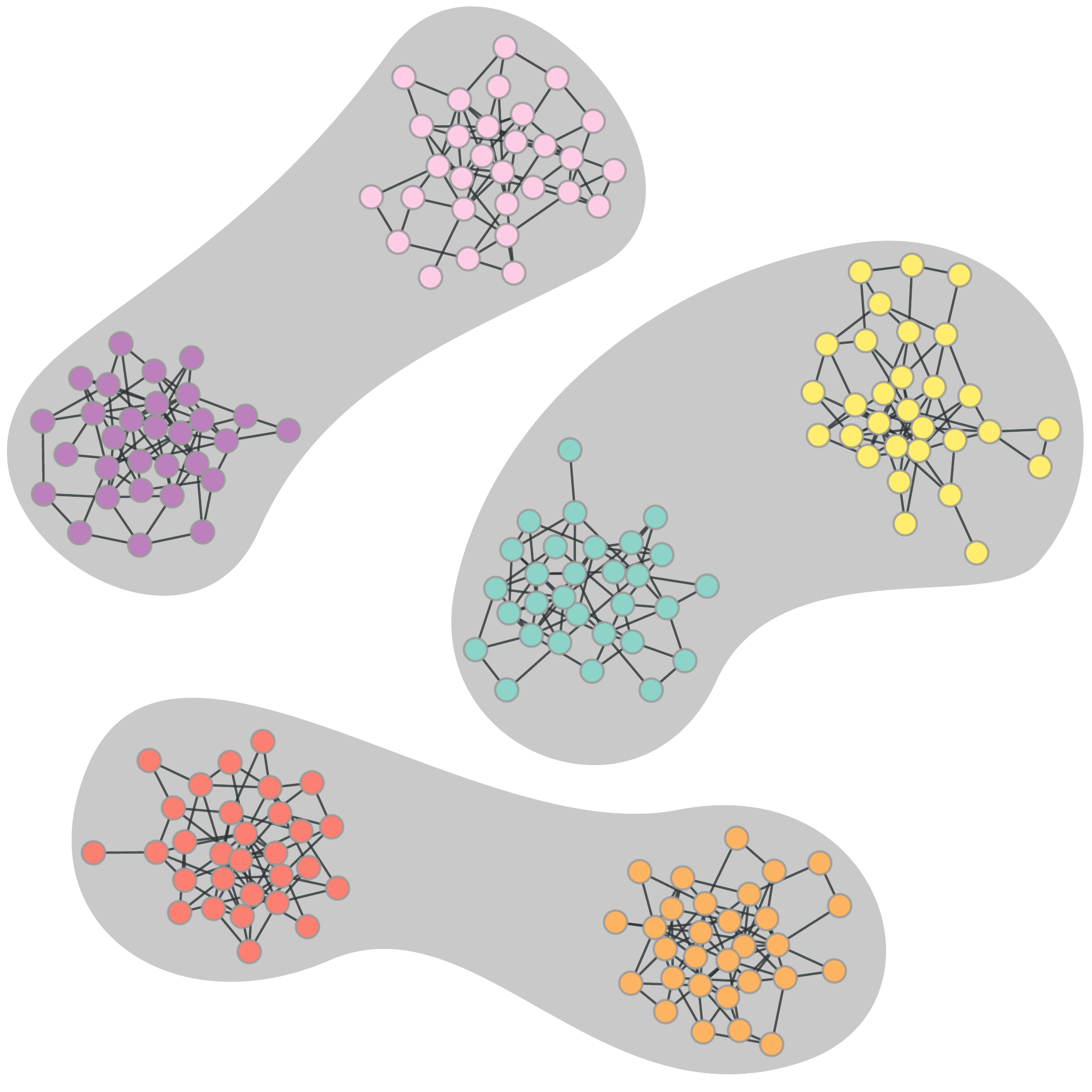} \caption{Planted
  partition of $B=6$ equal sized groups (node colors), being wrongly
  fitted as a $B^*=3$ model (shaded region). In the fitted model, the
  two groups inside each shaded region are not properly identified. This
  problem happens whenever $B > B_{\text{max}}$ for $B_{\text{max}} =
  O(\sqrt{N})$ using noninformative priors for the edge counts, but only
  for $B_{\text{max}} = O(N/\ln N)$ when the hierarchical priors are
  used instead.\label{fig:resolution}}
\end{figure}

We now pretend we have observed a network realization $\bm{A}$ sampled
from this ensemble, and compare the likelihood of the true partition
into $B$ groups with a wrong partition with $B^*<B$ groups. We will not
consider all possible wrong partitions; instead we will consider only
those where the correct planted groups were merged together into bigger
groups of equal size, as illustrated in Fig.~\ref{fig:resolution}. The
reason for this specific construction is twofold: 1. If we show that
this alternative partition has a higher likelihood than the plated one,
this would be sufficient to prove that the planted one will not be
detected by maximum likelihood;
2. The wrong fit induced by this alternative partition corresponds to
the original planted partition model that is only ``re-scaled'' by
replacing the planted value with the inferred one, $B\to B^*$, in
Eqs.~\ref{eq:pp_l} to~\ref{eq:pp_k}. This leaves us with a single
parameter to vary, allowing us to proceed with the analysis rather
easily. The inferred number of groups will be given simply by maximizing
the posterior likelihood
\begin{equation}
  B^* = \underset{q}{\operatorname{argmax}}\,\frac{P(\bm{A},\bm{k},\bm{e},\bm{b}(q))}{P(\bm{A})},
\end{equation}
where $\bm{b}(q) = \{\ceil{b_i q / B}\}$ is the re-scaled partition
according to parameter $q\in [1,B]$. Because of point 2 above, and as
long as $B^*\le B$, this amounts to maximizing the joint likelihood
given by Eqs.~\ref{eq:pp_l} to~\ref{eq:pp_k} with respect to the number
of groups $B$ replaced by $q$. If we assume that $N \gg 1$, $E \propto
N$, $B \gg 1$, as well as $N \gg B$ (although we make no assumption
between $B^2$ and $N$), and discard terms that do not depend on $B$, as
long as $q\le B$ we have
\begin{multline}\label{eq:asymp_l}
  \ln P(\bm{A},\bm{k},\bm{e},\bm{b}(q)) \approx (E - N)\ln q \\ - (E+q^2/2)h\left(\frac{E}{E+q^2/2}\right),
\end{multline}
where $h(x)=-x\ln x -(1-x)\ln(1-x)$.  If we maximize the above equation
with respect to $q$, we obtain
\begin{equation}
  B_{\text{max}} = x(\avg{k})\sqrt{N},
\end{equation}
with $x(\avg{k})$ being the solution of
\begin{equation}
  \avg{k} - 2 = 2x^2f'(1+x^2/\avg{k})
\end{equation}
with $f(x)=xh(1/x)$, and $\avg{k}=2E/N$. Since the re-scaling of the
likelihood is only valid for $B^* \le B$, we have that for any planted
partition with $B$ groups the actual inferred value will be $B^*=\min(B,
B_{\text{max}})$. Hence we obtain the same result of
Ref.~\cite{peixoto_parsimonious_2013} that the maximum number of groups
scales as $B_{\text{max}}\propto\sqrt{N}$. This property is robust with respect to
details of the model, and is simply a direct result of a noninformative
prior used for $P(\bm{e})$, which is responsible for the dependence on
$q^2$ in the last term of Eq.~\ref{eq:asymp_l}: A lack of prior
information on the large-scale structure incurs a cost in the
description length that scales roughly as $-\ln P(\bm{e}) \sim
(B^2/2)\ln E$ (for $B^2\gg E$). This means that we obtain very similar
results when considering the other model variants considered in this
work. In particular, using either Eq.~\ref{eq:canonical} or
\ref{eq:canonical_alt} we obtain asymptotic expressions for the joint
distribution that are very similar to Eq.~\ref{eq:asymp_l}, and yield
only a slightly worse scaling for the maximum number of groups,
$B_{\text{max}}\propto \sqrt{N/\ln N}$, with the $\sqrt{\ln N}$ difference due to
the priors of Eqs.~\ref{eq:lambda_prior_1} and ~\ref{eq:lambda_prior_2},
that allow the total number of edges to fluctuate. Using the uniform
hyperpriors for the degree sequences also has no effect on this
limitation.

On the other hand, as shown in Ref.~\cite{peixoto_hierarchical_2014},
this issue is significantly improved by using the hierarchical prior for
$\bm{e}$. Here we show this by considering a uniform hierarchical
division where at each level the number of groups decrease by a factor
$\sigma$, $B_l = B / \sigma^l$. Using Eq.~\ref{eq:prior_hierarchy}, we
have
\begin{multline}
  P(\bm{e}) = \prod_{l=1}^{\log_\sigma B}\multiset{\sigma(\sigma+1)/2}{2E\sigma^l/B}^{-B/\sigma^l}\times\\
  \frac{\sigma!^{B/\sigma^l}}{(B/\sigma^{l-1})!}{B/\sigma^{l-1} - 1\choose B/\sigma^l - 1}^{-1}.
\end{multline}
Assuming $B \gg \sigma$, and keeping only the leading terms, we have
$\ln P(\bm{e}) \approx -[B\sigma(\sigma+1) \ln E]/[2(\sigma-1)]$, and
hence
\begin{equation}
  \ln P(\bm{A},\bm{k},\bm{e},\bm{b}(q)) \approx (E - N)\ln q - \frac{\sigma(\sigma+1)}{2(\sigma-1)} q\ln E,
\end{equation}
from which we obtain the upper bound
\begin{equation}
  B_{\text{max}} = \frac{(\sigma-1)(\avg{k}-2)}{\sigma(\sigma+1)}\times \frac{N}{\ln N}.
\end{equation}
Hence, this choice of priors enables the identification of a number of
groups that is far larger that what is possible with the noninformative
choice. This comes with no drawbacks, since this prior includes the
noninformative one as a special case, and we are still protected against
overfitting; becoming only less susceptible to the \emph{underfitting}
that happens when $B > B_{\text{max}}$.

\section{Inference algorithm}\label{sec:mcmc}

The inference task we have is to sample from (or maximize) the posterior
distribution of the hierarchical partition,
\begin{equation}\label{eq:mcmc_posterior}
  P(\{\bm{b}_l\}|\bm{A}) =
  \frac{P(\bm{A},\{\bm{b}_l\})}{P(\bm{A})}.
\end{equation}
The approach we will take is based on a Markov chain Monte Carlo
importance sampling for the partitions at all hierarchy levels. The
algorithm will revolve around moving the membership of nodes in
different hierarchical levels at random, and accepting or rejecting
those moves, so that after a sufficiently long equilibration time, the
hierarchical partitions are sampled according to
Eq.~\ref{eq:mcmc_posterior}. We note that this posterior can be
factorized as
\begin{align}
  P(\{\bm{b}_l\}|\bm{A}) &= \frac{\prod_lP(\bm{e}_{l-1},\bm{b}_l|\bm{e}_l)}{P(\bm{A})}\nonumber\\
  &=\prod_lP(\bm{b}_l|\bm{e}_{l-1},\bm{e}_l)
\end{align}
with per-level posteriors
\begin{align}\label{eq:l_posterior}
  P(\bm{b}_l|\bm{e}_l,\bm{e}_{l+1}) &= \frac{P(\bm{e}_l|\bm{e}_{l+1},\bm{b}_l)P(\bm{b}_l)}{P(\bm{e}_l|\bm{e}_{l+1})},
\end{align}
where we assume $\bm{e}_0=A$, and $P(\bm{e}_l|\bm{e}_{l+1})$ is a
normalization constant.

Therefore, a workable approach is to separately sample partitions at
each level according to its individual posterior, conditioned on the
remaining levels, which are kept unchanged for the time being. If we
sample from each level in this manner we can guarantee ergodicity, and
if the moves at the individual levels are reversible, the overall
distribution will correspond to the desired full posterior of
Eq.~\ref{eq:mcmc_posterior}. Since the hierarchical levels are coupled,
when moving a node at level $l$, we must ensure that this does not
invalidate the partition at level $l+1$. Hence, we must forbid node
moves between groups that are themselves at different groups in the next
level. (This constraint does not break ergodicity, since all partitions
in the upper levels will be allowed to change at some point).

In more detail, we proceed as follows. At each individual level $l$, we
perform a move proposal of node $i$ from its current group $r$ to a new
group $s$, according to a probability $P(b_i^{(l)}=r\to s)$ that we will
specify shortly. We compute the difference in the log-likelihood
$\Delta\ln P_l$ at that level, and we accept the move according to the
Metropolis-Hastings
criterion~\cite{metropolis_equation_1953,hastings_monte_1970}, i.e. with
a probability
\begin{equation}\label{eq:metropolis}
  a = \min\left\{1, e^{\Delta\ln P_l}\frac{P(b_i^{(l)}=s\to r)}{P(b_i^{(l)}=r\to s)}\right\},
\end{equation}
where $P(b_i^{(l)}=s\to r)$ is the probability of the reverse move being
proposed. The log-likelihood difference is computed as
\begin{equation}\label{eq:delta_L}
  \Delta\ln P_l = \ln \frac{P(b_i^{(l)}=s,\bm{b}_l\smallsetminus b_i^{(l)}|\bm{e}_l,\bm{e}_{l+1})}
            {P(b_i^{(l)}=r,\bm{b}_l\smallsetminus b_i^{(l)}|\bm{e}_l,\bm{e}_{l+1})},
\end{equation}
where $\bm{b}_l\smallsetminus b_i^{(l)}$ means the partition of the
remaining nodes excluding node $i$. Note that in computing
Eq.~\ref{eq:delta_L}, we do not need to determine the normalization
constant in Eq.~\ref{eq:l_posterior}, and the remaining relevant terms
correspond only to a subset of the full joint distribution of
Eq.~\ref{eq:full_joint_big}. Typically, the number of groups in the
upper levels decreases exponentially, and hence the algorithmic
complexity is dominated by the bottom level $l=0$. As mentioned
previously, the number of terms of the joint distribution that are
necessary to compute $\Delta\ln P_0$ is proportional only to the degree
$k_i$ of node $i$, and is independent of $B_1$, and hence can be
computed quickly. Therefore, if we attempt one move for each node in the
network, such a ``sweep'' can be completed in time $O(E)$, independent
on the total number of groups.

An important element of this algorithm is the move proposal probability
$P(b_i^{(l)}=r\to s)$.  Any choice with nonzero probability for all
values of $s$ will preserve ergodicity, and --- coupled with the
Metropolis-Hastings criterion --- also detailed balance. These two
ingredients are sufficient to guarantee that hierarchical partitions are
eventually sampled from the correct posterior distribution. However, in
practice, the equilibration time will depend strongly on the move
proposals, and will become shorter if they are close to the actual
posterior. The simplest choice we could make is to select from all
groups with equal probability
\begin{equation}
  P(b_i^{(l)}=r\to s) = \frac{1}{B_l+1},
\end{equation}
where we also account for the occupation of a new group, which if the
move is accepted, will increase $B_l$ by one (provided the node $i$ is
not the last one in its current group). Since this probability is always
nonzero, it fulfills our requirements. However, it will lead to very
large equilibration times, in particular for large values of $B_l$. This
is because the actual posterior distribution for node $i$ is likely to
be concentrated only in a small subset of all possible groups, and hence
most such fully random proposals will simply be rejected. A better
approach was developed in Ref.~\cite{peixoto_efficient_2014}, and it
consists in inspecting the current parameters of the model to provide a
better guess of the posterior. It amounts to making move proposals
according to
\begin{equation}
  P(b_i^{(l)}=r\to s) = \sum_tP(t|i,l)\frac{e_{ts}^l + \epsilon}{e_t^l + \epsilon(B_l+1)},
\end{equation}
where $P(t|i,l) = \sum_jA_{ij}^{(l)}\delta({b_j^{(l)},t})/k_i^{(l)}$ is the
fraction of neighbors of node $i$ in level $l$ that belong to group $t$,
and $\epsilon > 0$ is an arbitrary parameter that enforces ergodicity,
but with no other significant impact in the algorithm, provided it is
sufficiently small. It is worthwhile to emphasize that these move
proposals do not bias the partitions toward any particular mixing
pattern. For example, they do not prefer assortative versus
non-assortative partitions, since they inspect the neighbors of a node
only to access with other groups their kinds are typically connected ---
which can be different from the the group assignment of the original
node. Furthermore, these proposals can be generated efficiently, simply
by
\begin{enumerate}
  \item sampling a random neighbor $j$ of
    node $i$, and inspecting its group membership $t=b_j$, and then
  \item with probability $\epsilon(B_l+1)/(e_t + \epsilon(B_l+1))$ sampling
        a fully random group $s$ (which can be a new group),
  \item or otherwise, sampling a group label $s$ with a
    probability proportional to the number of edges leading to it from
    group $t$, $e_{ts}$.
\end{enumerate}
The above can be done in time $O(k_i)$, again independently of $B_l$, as
long as a continuous book-keeping is made of the edges which are
incident to each group, and therefore it does not affect the overall
$O(E)$ time complexity. As reported in
Ref.~\cite{peixoto_efficient_2014}, these move proposals tend to
significantly improve the mixing times, and remove an explicit
dependency on the number of groups, that would otherwise be present with
the fully random moves.

This approach is also more efficient than the rejection-free ``heat
bath'' algorithm used in Ref.~\cite{newman_estimating_2016}, since the
latter requires all possible moves to be probed, incurring an additional
time complexity that grows linearly with the number of groups.

In addition to the move proposals, another crucial aspect of the
algorithm's efficiency is the choice of the starting state. A simple
approach such as starting from a random partition can lead to metastable
states, from which it takes a long time to escape. Instead, here we
adopt the agglomerative initialization approach presented in
Ref.~\cite{peixoto_efficient_2014}, which amounts to putting each node
in their own group, and then progressively merging groups, while
alternatingly allowing for individual node moves. This can be done for
each hierarchical level iteratively, as described in detail in
Ref.~\cite{peixoto_hierarchical_2014}. As reported in
Ref.~\cite{peixoto_efficient_2014}, this approach greatly reduces the
tendency to get trapped in a metastable state, and serves as an
initialization protocol that further reduces the overall mixing time of
the MCMC.

While the above algorithm serves to sample from the posterior
distribution of Eq.~\ref{eq:mcmc_posterior}, it can be easily modified
to find its maximum by introducing an ``inverse-temperature'' parameter
$\beta$ in Eq.~\ref{eq:metropolis} via the replacement $\Delta\ln
P_l\to\beta\Delta\ln P_l$. By making $\beta\to\infty$ the algorithm is
turned into a greedy heuristic that, if repeated many times, yields a
reliable estimate of the maximum.

The lack of an explicit dependence on the number of groups of the
algorithm above is atypical, since most other proposed Bayesian (or
semi-Bayesian) algorithms have either quadratic
$O(EB^2)$~\cite{yan_active_2010,decelle_inference_2011,
come_model_2015,newman_estimating_2016} or linear
$O(EB)$~\cite{guimera_missing_2009,gopalan_efficient_2013} dependencies,
which means that those can be applied to large networks only if the
number of groups is kept small. Furthermore, the increased efficiency
obtained here does not rely on any approximations made to the
likelihood.

A reference implementation of the algorithm is freely available as part
of the \texttt{graph-tool}
library~\cite{peixoto_graph-tool_2014}\footnote{Available
at~\url{https://graph-tool.skewed.de}.}.

\section{Model comparison}\label{sec:comparison}

With the three different model flavors available (NDC-SBM, DC-SBM with
uniform degree prior or uniform hyperprior) we are left with the problem
of deciding which offers the best description of a given network. This
problem can be formulated in at least two ways, depending on whether we
want to compare individual partitions or entire model classes, which we
describe now detail.

If we wish to compare two individual partitions, obtained from the
posterior distribution of two different models, wee need to consider the
joint posterior probability $P(\{\bm{b}_l\}, \mathcal{H}|\bm{A})$, where
$\mathcal{H}$ is the model class being used. For example, when comparing
results from the DC-SBM and NDC-SBM, we can compute the ratio,
\begin{align}\label{eq:lambda1}
   \Lambda_1 &= \frac{P(\{\bm{b}_l\}, \mathcal{H}_\text{NDC} | \bm{A})}{P(\{\bm{b}_l\}', \mathcal{H}_\text{DC} | \bm{A})} \nonumber\\
           &= \frac{P(\bm{A}, \{\bm{b}_l\} | \mathcal{H}_\text{NDC})}{P(\bm{A}, \{\bm{b}_l\}' | \mathcal{H}_\text{DC})}\times\frac{P(\mathcal{H}_\text{NDC})}{P(\mathcal{H}_\text{DC})} \nonumber\\
           &= 2^{-\Delta\Sigma}
\end{align}
where in the last equation $\Delta\Sigma = \Sigma_\text{NDC} -
\Sigma_\text{DC}$ is the difference in the description length, and we
have assumed that both model classes are equally likely a priori,
$P(\mathcal{H}_\text{NDC})=P(\mathcal{H}_\text{DC})$. If $\Lambda_1 <
1$, we have that the data favors the particular hierarchical partition
$\{\bm{b}_l\}'$ together with the degree-corrected model variant, or if
$\Lambda_1 > 1$ we have the opposite case. Choosing a model according to
$\Lambda_1$ is identical to employing the MDL criterion, but its value
can be used to quantify the degree of confidence. E.g. a value
$\Lambda_1 = 1/2$ indicates a very modest evidence supporting the DC-SBM
that cannot be reliably distinguished from pure chance, whereas a value
of $\Lambda_1 = 1/10^5$ would clearly indicate that it is a much better
model than the NDC-SBM.

The criterion above should not be confused with the ``frequentist''
approach of computing the \emph{parametric} likelihood ratio between
both models, as was done in Ref.~\cite{yan_model_2014}. In the latter
case, which does not involve any prior probabilities, the ratio needs to
be compared to the distribution obtained with the null model, which is
more cumbersome to obtain. However, as is understood in general (and can
also be shown for the particular case of the
SBM~\cite{yan_bayesian_2016}), this frequentist criterion should
coincide asymptotically with the Bayesian criterion above as long as
uniform priors are used. On the other hand, since here we use deeper
Bayesian hierarchies, and hence nonuniform priors, these amount to
different tests, with $\Lambda_1$ being more sensitive to regularities
in the data, since it uses properties of the parameters themselves in
the decision.

The comparison above using $\Lambda_1$ is easy to perform, since it
requires one to simply inspect the result of the inference
procedure. However, it may be possible that the same network admits many
alternative fits with very similar posterior probabilities. A more
strict Bayesian stance would require us to treat those on an equal
footing, and any statement about the generative model behind the data
should be averaged over all possible fits, weighted according to the
respective posterior probability. Hence, in this scenario we may be
interested instead in comparing the entire model classes to each other,
which involves evaluating the so-called \emph{model evidence} by summing
over all hierarchical partitions,
\begin{equation}\label{eq:evidence_h}
  P(\bm{A}|\mathcal{H}) = \sum_{\{\bm{b}_l\}}P(\bm{A},\{\bm{b}_l\}).
\end{equation}
With this, we can again compute the posterior odds ratio, e.g.
\begin{equation}\label{eq:lambda2}
   \Lambda_2 = \frac{P(\mathcal{H}_\text{NDC} | \bm{A})}{P(\mathcal{H}_\text{DC} | \bm{A})} 
   = \frac{P(\bm{A} | \mathcal{H}_\text{NDC})}{P(\bm{A} | \mathcal{H}_\text{DC})}\times\frac{P(\mathcal{H}_\text{NDC})}{P(\mathcal{H}_\text{DC})}.
\end{equation}
If we have no prior preference towards either
model, $P(\mathcal{H}_\text{NDC})=P(\mathcal{H}_\text{DC})$, the value
of $\Lambda_2$ is known as the Bayes factor~\cite{jeffreys_theory_1998},
and like $\Lambda_1$ can be used to establish a degree of confidence in
the outcome.

Unfortunately, the exact computation of the sum in
Eq.~\ref{eq:evidence_h} is intractable. We therefore resort to a
variational approach, firstly by writing
\begin{align}\label{eq:free-energy}
  \ln P(\bm{A}|\mathcal{H}) &= \ln \sum_{\{\bm{b}_l\}}P(\bm{A},\{\bm{b}_l\})\\
   &=\sum_{\{\bm{b}_l\}}q({\{\bm{b}_l\}})\ln P(\bm{A},\{\bm{b}_l\})\label{eq:avg_logp}\\
   &\qquad\qquad-\sum_{\{\bm{b}_l\}}q({\{\bm{b}_l\}})\ln q({\{\bm{b}_l\}})\label{eq:q_ent},
\end{align}
with
\begin{equation}
  q({\{\bm{b}_l\}}) = \frac{P(\bm{A},\{\bm{b}_l\})}{P(\bm{A})}
\end{equation}
being precisely the posterior distribution of for the hierarchical
partition that we obtain from with the MCMC algorithm used above. (Note
that so far we have not made any approximations, with the identities
above holding exactly.) The first term in Eq.~\ref{eq:avg_logp} is easy
to compute, as it amounts to the average log-likelihood (or minus the
description length) of the partitions we obtain with the MCMC above,
\begin{equation}
  \avg{\ln P(\bm{A},\{\bm{b}_l\})} = \sum_{\{\bm{b}_l\}}q({\{\bm{b}_l\}})\ln P(\bm{A},\{\bm{b}_l\}).
\end{equation}
On the other hand, the second term in Eq.~\ref{eq:q_ent} amounts to the
entropy of the posterior distribution,
\begin{equation}
  H({\{\bm{b}_l\}}) = -\sum_{\{\bm{b}_l\}}q({\{\bm{b}_l\}})\ln q({\{\bm{b}_l\}}),
\end{equation}
and measures how strongly it is concentrated. For example, in the
extreme (and unrealistic) case where for each model being compared only
one partition occurs with probability $q({\{\bm{b}_l\}})=1$, the entropy
will be zero, and we have that $\Lambda_1 = \Lambda_2$. Otherwise the
entropy $H({\{\bm{b}_l\}})$ will effectively measure how many partitions
contribute to the average log-likelihood, so that a model class with a
larger entropy will be preferred over another with less variance, even
if their posterior probabilities are on average the same. Unfortunately,
the entropy $H({\{\bm{b}_l\}})$ is notoriously difficult to compute
exactly, even asymptotically via MCMC algorithms, and encapsulates the
difficulty of computing Eq.~\ref{eq:evidence_h} directly. A brute force
approach simply does not work, since it would require keeping track of
all visited hierarchical partitions, which grow combinatorially in
number with system size. Other approaches such as thermodynamic
integration~\cite{frenkel_understanding_2001}, annealed importance
sampling~\cite{neal_annealed_2001} and flat-histogram
methods~\cite{wang_efficient_2001} are also possible, but tend to be
significantly inefficient in comparison. Instead, here we make a
so-called ``mean field'' assumption on the shape of $q({\{\bm{b}_l\}})$
which assumes that it factorizes over all levels
\begin{equation}\label{eq:mf}
  q({\{\bm{b}_l\}}) \approx q_i^1(\bm{b}_1)\prod_{l>1}\prod_iq_i^l(b_i^l).
\end{equation}
For the first level we use the so-called ``Bethe
approximation''~\cite{mezard_information_2009}, which takes into account
the correlation between adjacent nodes in the network,
\begin{equation}
   q^1(\bm{b}_1) \approx \prod_{i<j}\left[q_{ij}^1(b_i^1,b_j^1)\right]^{A_{ij}}\prod_i\left[q_i^1(b_i^1)\right]^{1-k_i}
\end{equation}
with $q^1_i(r)$ and $q_{ij}^1(r,s)$ obtained from the posterior node and
edge marginals
\begin{align}
  q^l_i(r) &= P(b_i^l=r|\bm{A}) = \sum_{\{\bm{b}^l\}\smallsetminus b_i^l}P(b_i^l=r,\{\bm{b}^l\}\smallsetminus b_i^l|\bm{A}), \\
  q_{ij}^1(r, s) &= P(b_i^1=r,b_j^1=s|\bm{A}) \nonumber\\
  &=\sum_{\{\bm{b}^l\}\smallsetminus \{b_i^1,b_j^1\}}P(b_i^1=r,b_j^1=s,\{\bm{b}^l\}\smallsetminus \{b_i^1,b_j^1\}|\bm{A}),
\end{align}
estimated with the MCMC algorithm above. For the upper levels $l>1$ we
cannot use the same approximation since the adjacency matrices will be
in general multigraphs that will keep changing throughout the
algorithm. Therefore we used above a mean-field approximation where the
posterior factorizes over all nodes. With this we can finally write
Eq.~\ref{eq:free-energy} as
\begin{equation}\label{eq:free-energy-approx}
\ln P(\bm{A}) \approx \avg{\ln P(\bm{A},\{\bm{b}_l\})} + \sum_lH_l
\end{equation}
where
\begin{multline}
  H_1 = -\sum_{i<j}A_{ij}\sum_{rs}q_{ij}^1(r,s)\ln q_{ij}^1(r,s) \\
  - \sum_i(1-k_i)\sum_rq_i^1(r)\ln q_i^1(r)
\end{multline}
is the entropy of the first level and
\begin{equation}
  H_l = -\sum_i\sum_rq^l_i(r)\ln q^l_i(r)
\end{equation}
is the entropy of the remaining hierarchy levels $l>1$. Thus,
Eq.~\ref{eq:free-energy-approx} can be computed simply by equilibrating
the MCMC, obtaining the average log-likelihood and the node and edge
posterior marginal distribution, $q^l_i(r)$ and $q^1_{ij}(r,s)$.

\section{Results for empirical networks}\label{sec:empirical}

\begin{figure*}
  \begin{tabular}{CCC}\smaller
    \includegraphics[width=.32\textwidth]{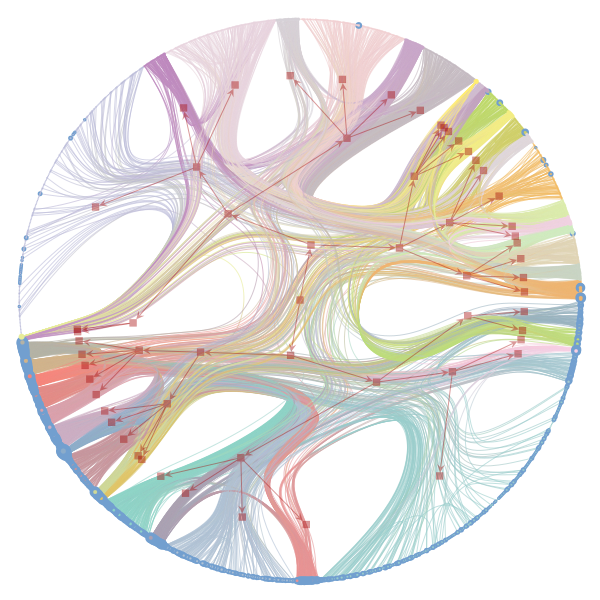}&
    \includegraphics[width=.32\textwidth]{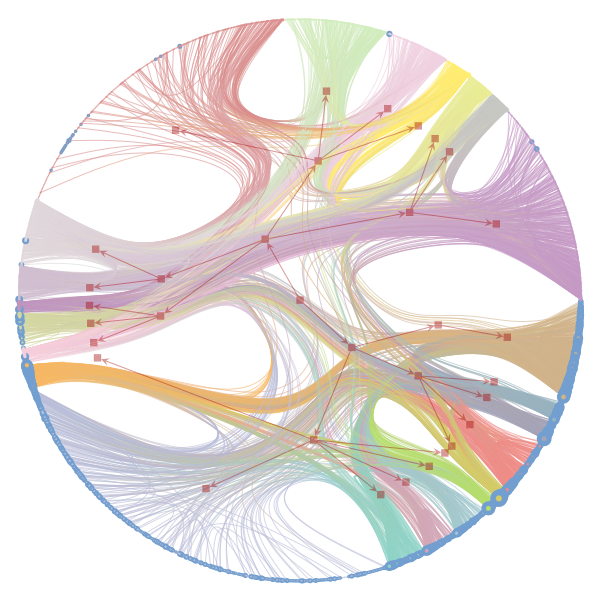}&
    \includegraphics[width=.32\textwidth]{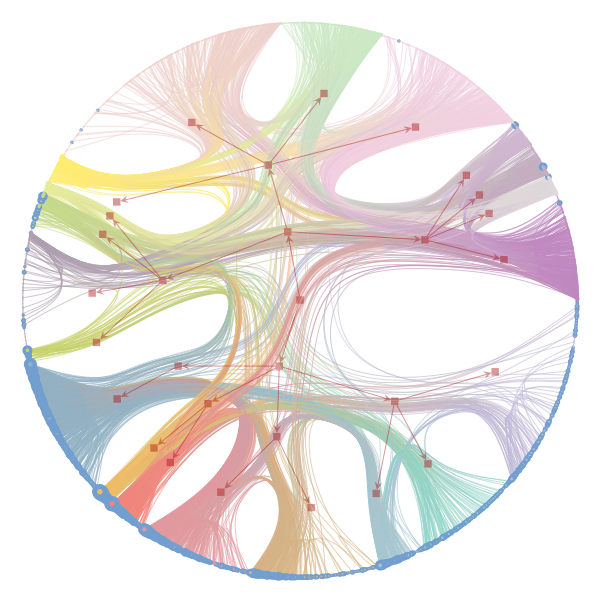}\\
    (a)&
    (b)&
    (c)
  \end{tabular}

  \caption{Most likely hierarchical partitions of a network of political
  blogs~\cite{adamic_political_2005}, according to the three model
  variants considered, as well as the number of groups $B_1$ at the
  bottom of the hierarchy, and the description length $\Sigma$:
  (a) NDC-SBM, $B_1=42$, $\Sigma \approx 89938$ bits,
  (b) DC-SBM, uniform prior, $B_1=23$, $\Sigma \approx 87162$ bits,
  (c) DC-SBM, uniform hyperprior, $B_1=20$, $\Sigma \approx 84890$ bits.
  The nodes circled in blue were classified as ``liberals'' and the
  remaining ones as ``conservatives'' in
  Ref.~\cite{adamic_political_2005} based on the blog contents.  Note
  that in all cases this division in two groups is correctly identified
  at the topmost level of the hierarchy. However, the lower levels yield
  significantly different subdivisions depending on which model type is
  used. The layout is obtained with an algorithm by
  Holten~\cite{holten_hierarchical_2006}. \label{fig:polblogs}}
\end{figure*}

\begin{figure*}
  \begin{minipage}{.36\textwidth}
    \begin{overpic}[width=\textwidth]{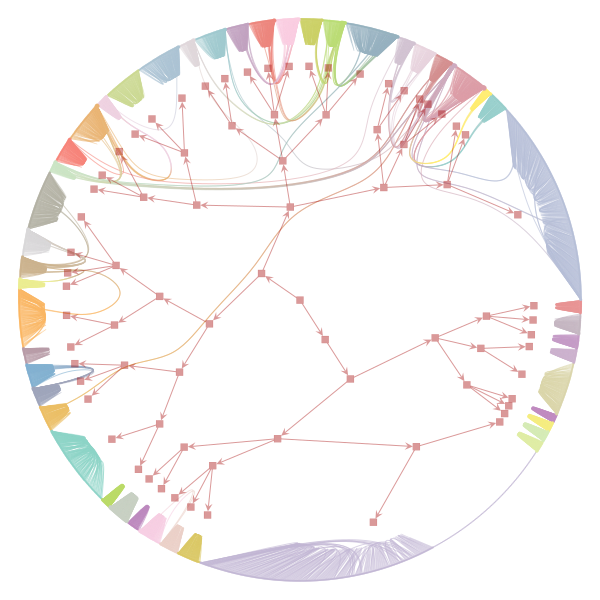}
      \put(0,0){\smaller(a)}
    \end{overpic}
  \end{minipage}
  \begin{minipage}{.52\textwidth}
    \begin{tabular}{ccc}
    \begin{overpic}[width=.33\textwidth]{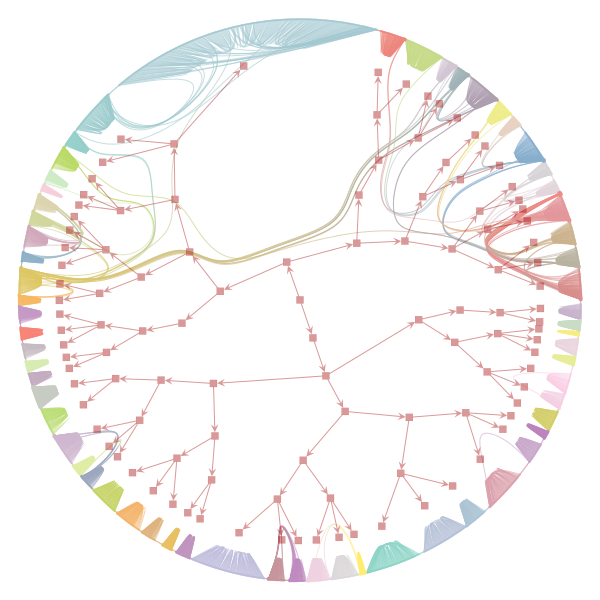}
    \end{overpic}&
    \begin{overpic}[width=.33\textwidth]{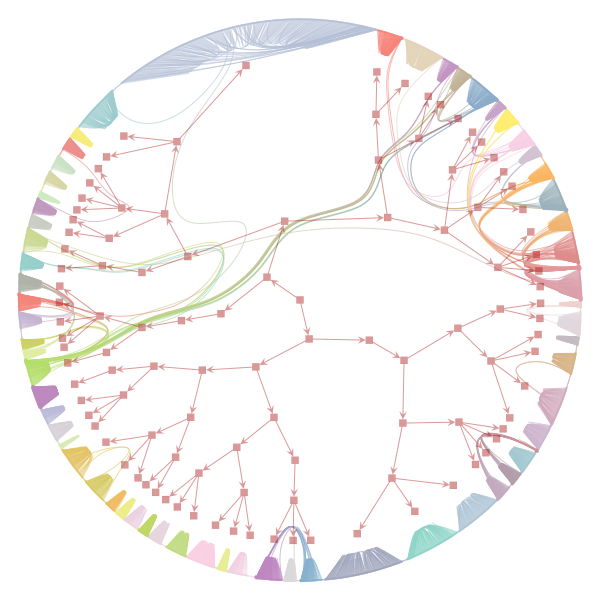}
    \end{overpic}&
    \begin{overpic}[width=.33\textwidth]{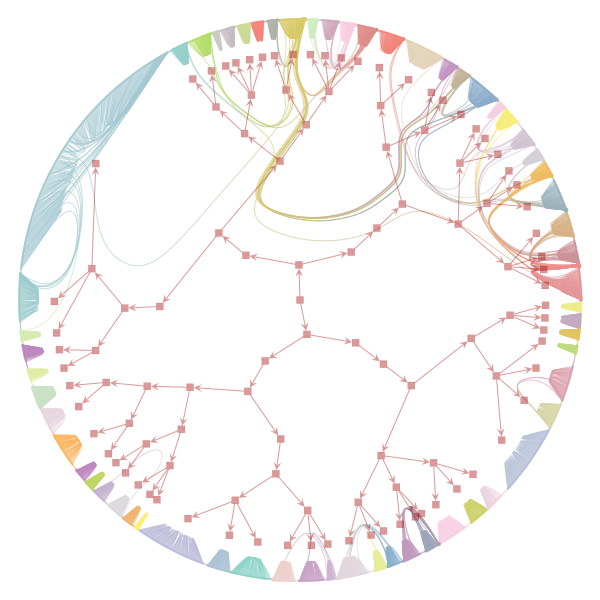}
    \end{overpic}\\
    \begin{overpic}[width=.33\textwidth]{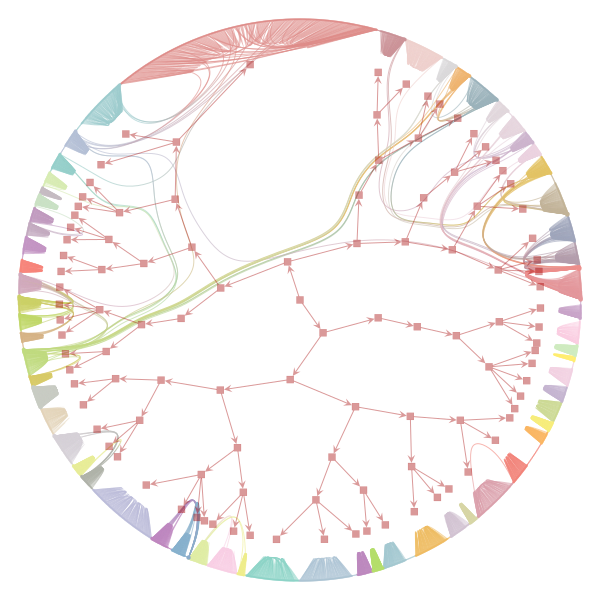}
      \put(0,0){\smaller(b)}
    \end{overpic}&
    \begin{overpic}[width=.33\textwidth]{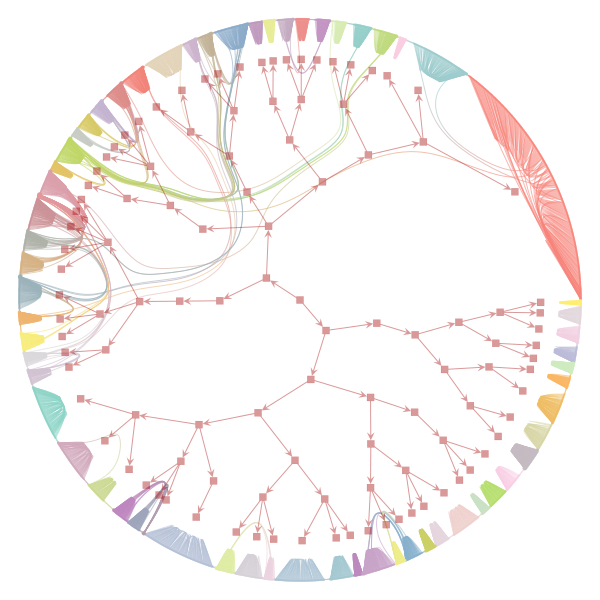}
    \end{overpic}&
    \begin{overpic}[width=.33\textwidth]{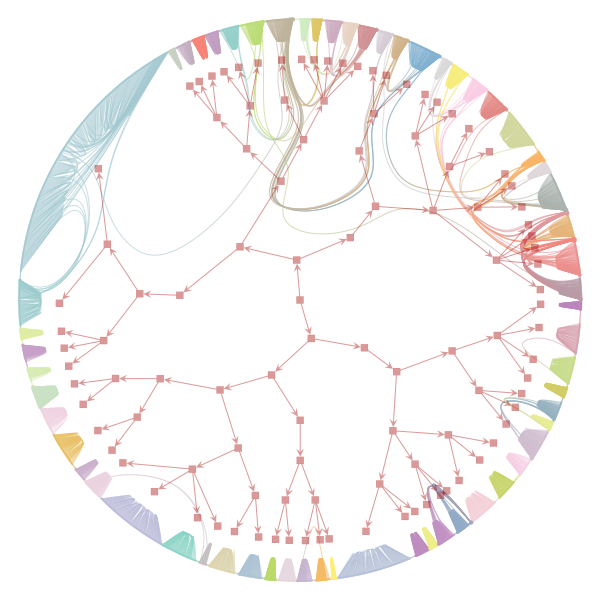}
    \end{overpic}
    \end{tabular}
  \end{minipage}
  \begin{minipage}{\textwidth}
    \includegraphics[width=\textwidth]{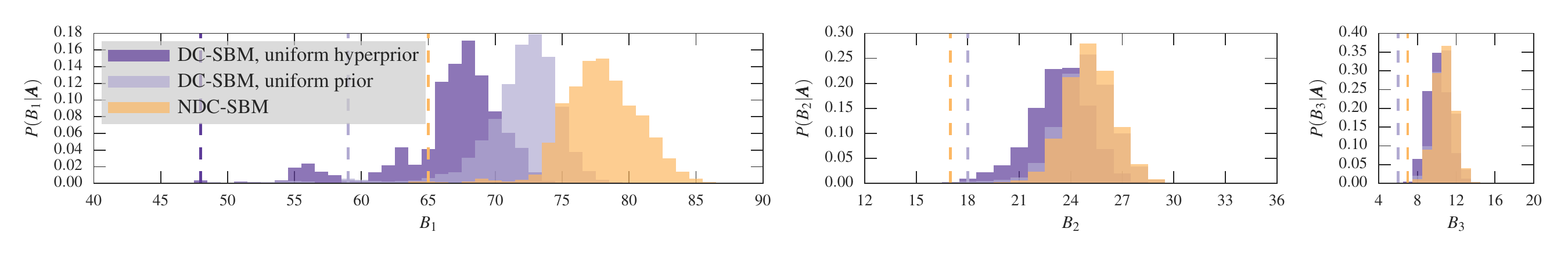}
  \end{minipage}

  \caption{Hierarchical partitions of a network of collaboration between
  scientists~\cite{newman_finding_2006}. (a) Most likely hierarchical
  partition according to the DC-SBM with a uniform hyperprior. (b)
  Uncorrelated samples from the posterior distribution. (c) Marginal
  posterior distribution of the number of groups at the first three
  hierarchical levels, according to the model variants described in the
  legend. The vertical lines mark the value obtained for the most likely
  partition.\label{fig:netscience}}
\end{figure*}

\begin{figure*}
  \begin{minipage}{.49\textwidth}
    (a)\\
    \includegraphics[width=\textwidth]{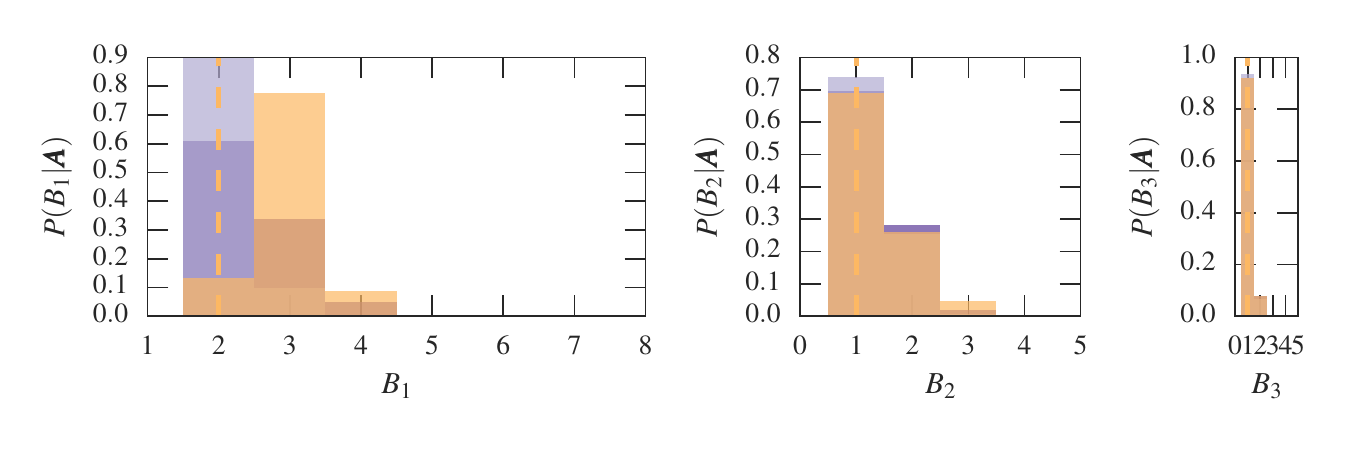}
  \end{minipage}
  \begin{minipage}{.49\textwidth}
    (b)\\
    \includegraphics[width=\textwidth]{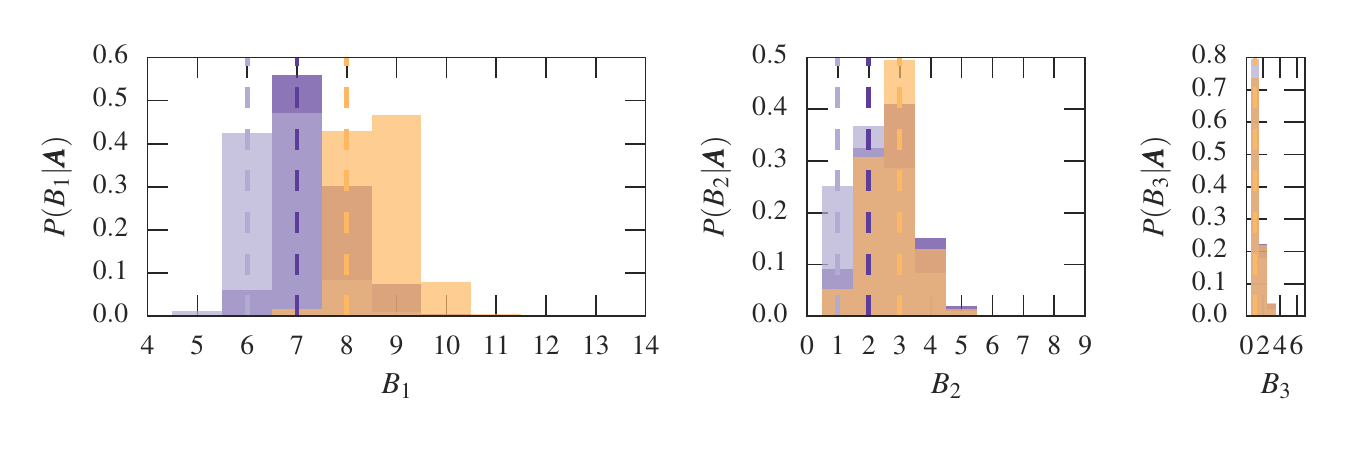}
  \end{minipage}
  \begin{minipage}{.49\textwidth}
    (c)\\
    \includegraphics[width=\textwidth]{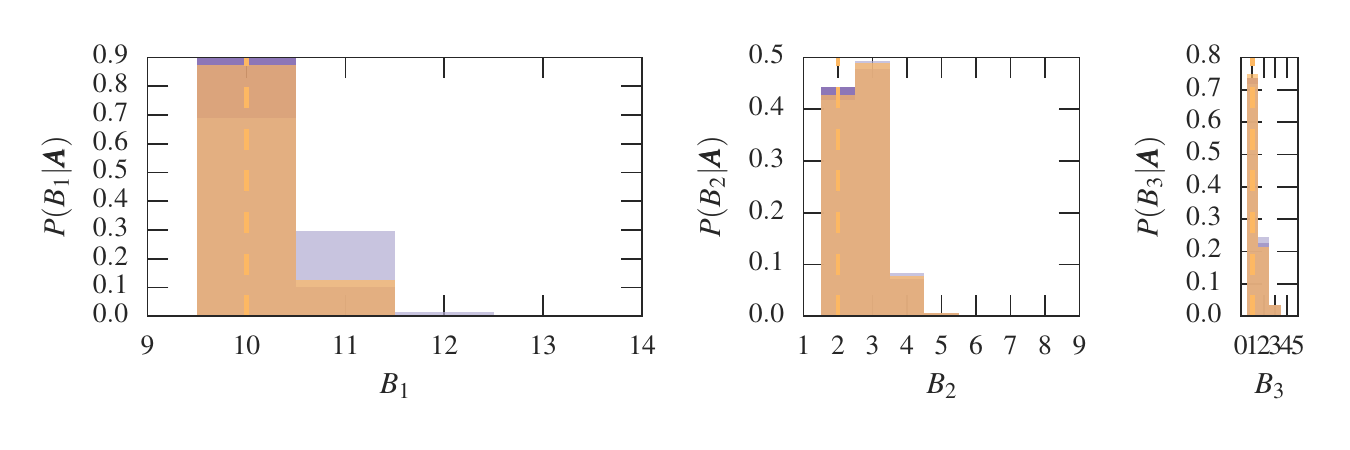}
  \end{minipage}
  \begin{minipage}{.49\textwidth}
    (d)\\
    \includegraphics[width=\textwidth]{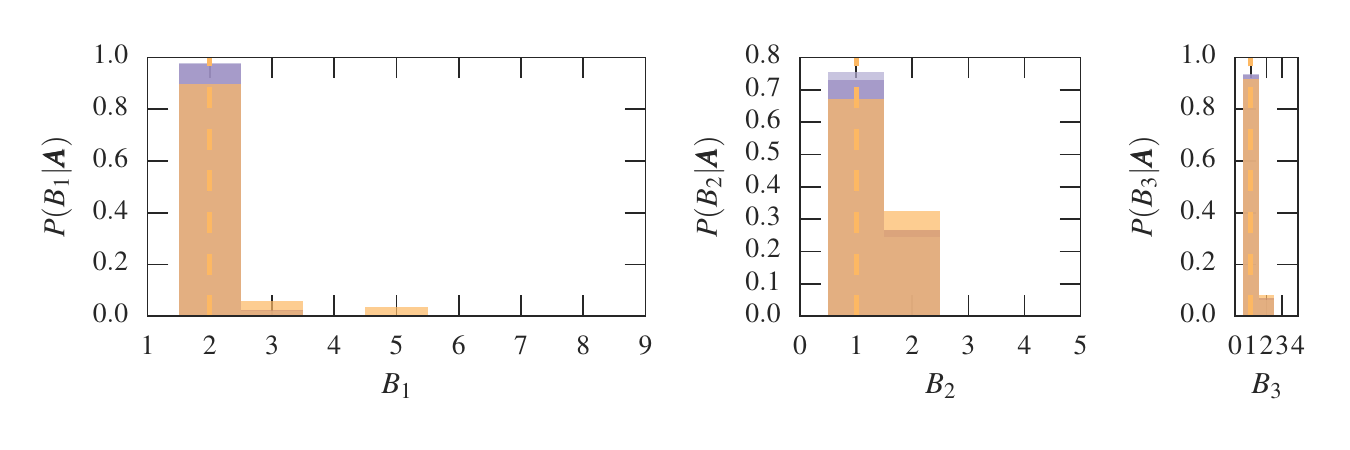}
  \end{minipage}
  \begin{minipage}{\textwidth}
    (e)\\
    \includegraphics[width=\textwidth]{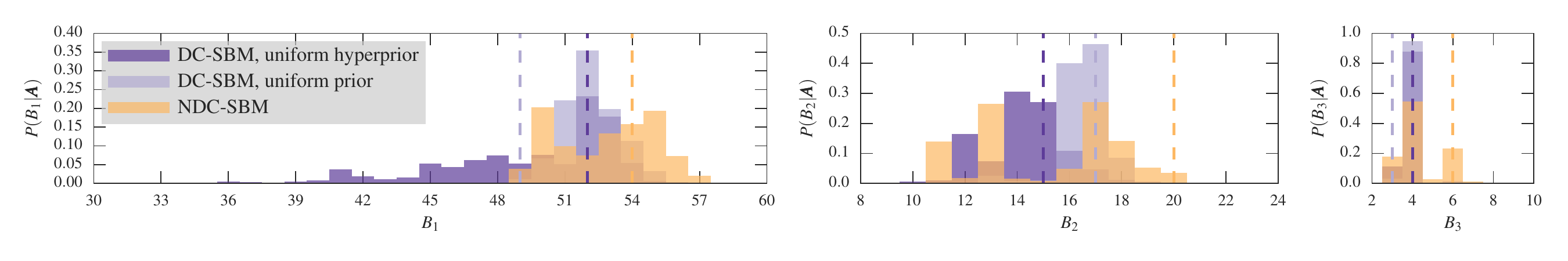}
  \end{minipage}
  \begin{minipage}{\textwidth}
    (f)\\
    \includegraphics[width=\textwidth]{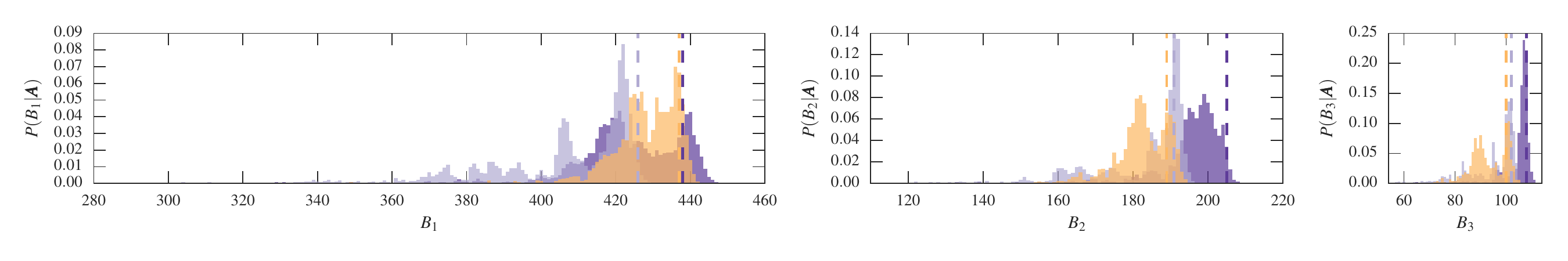}
  \end{minipage}
  \begin{minipage}{\textwidth}
    (g)\\
    \includegraphics[width=\textwidth]{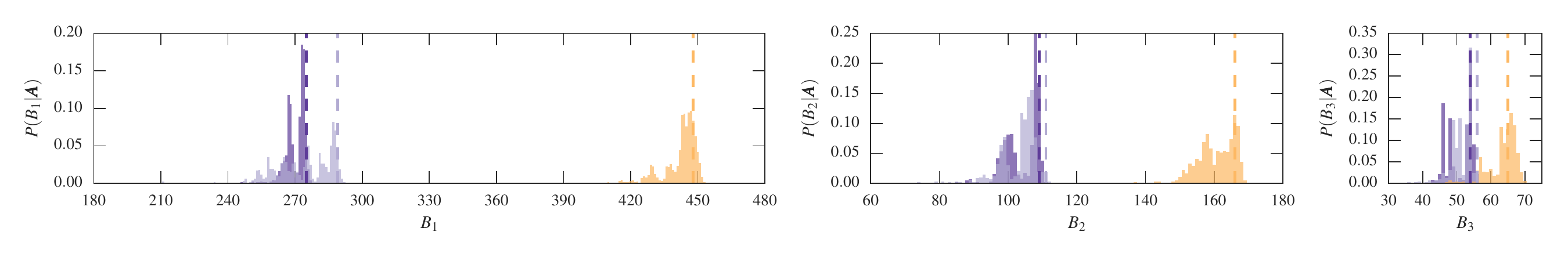}
  \end{minipage}
  \begin{minipage}{\textwidth}
    (h)\\
    \includegraphics[width=\textwidth]{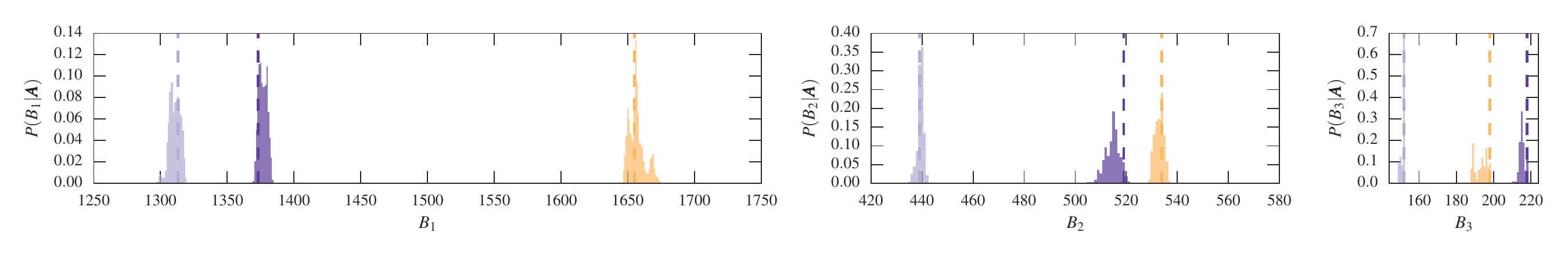}
  \end{minipage}

  \caption{Marginal posterior distribution of the number of groups at
  the first three hierarchical levels, according to the model variants
  described in the legend, for some of the empirical networks listed in
  table~\ref{tab:data}:
  (a) Dolphin social network,
  (b) Characters in Les Misérables,
  (c) American college football,
  (d) Southern women interactions,
  (e) Malaria gene similarity,
  (f) Protein interactions (II),
  (g) Global airport network,
  (h) Dictionary entries.
  The vertical
  lines mark the value obtained for the most likely partition (the MDL
  criterion). \label{fig:B-posterior}}
\end{figure*}

\begin{figure*}
  \begin{overpic}[width=\textwidth]{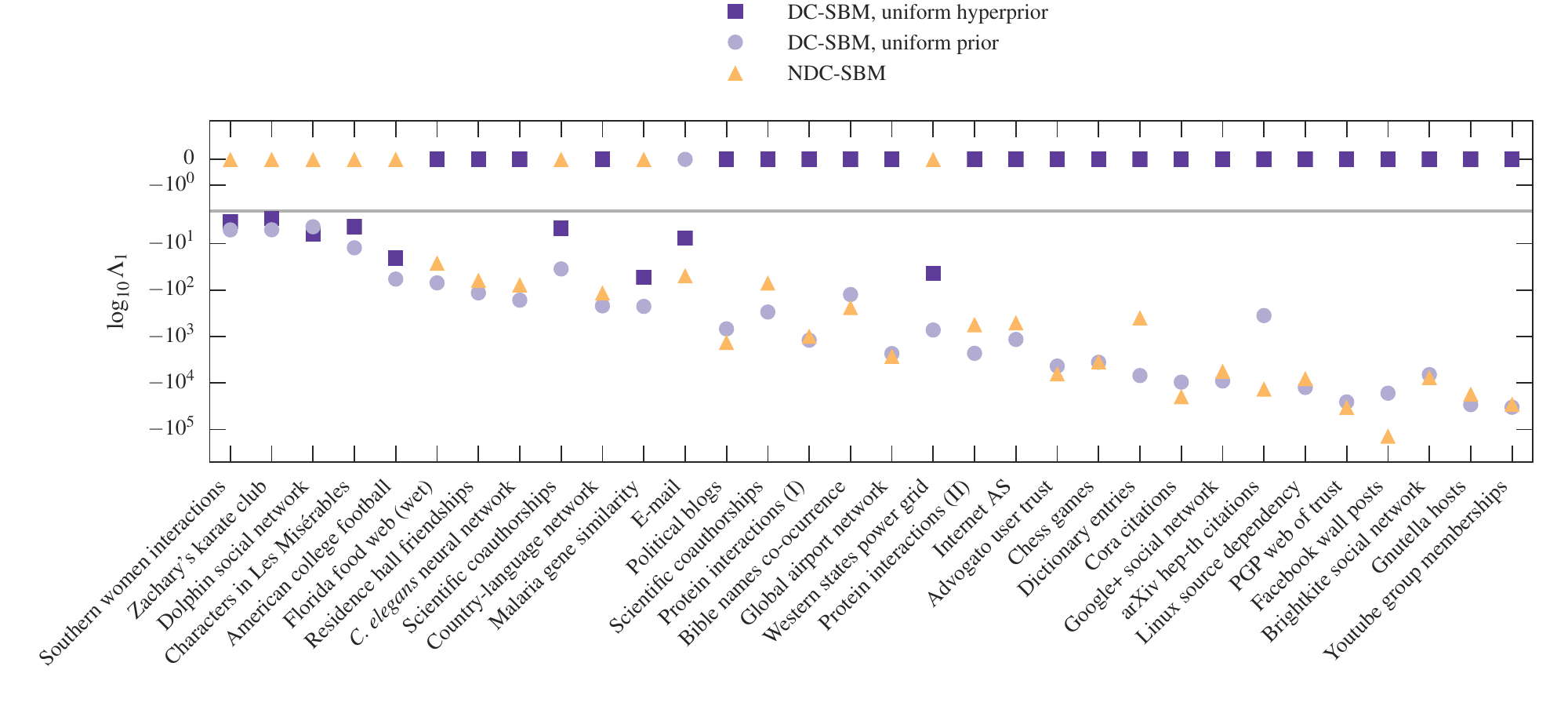}
    \put(0,36){\smaller(a)}
  \end{overpic}
  \begin{overpic}[width=\textwidth, trim={0 0 0 1.2cm}, clip]{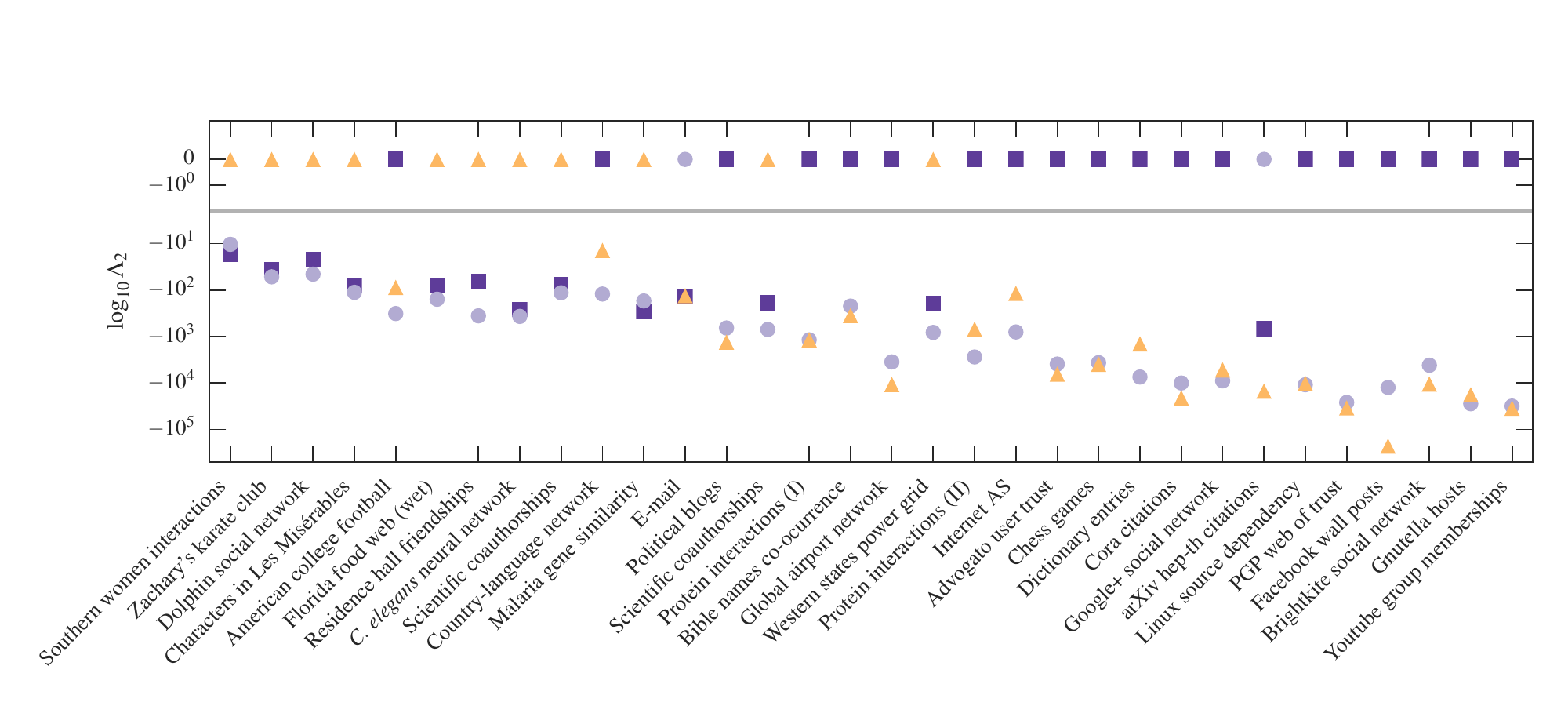}
    \put(0,36){\smaller(b)}
  \end{overpic}

  \caption{Posterior odds ratio relative to the best model, according to
  (a) the MDL criterion, $\Lambda_1$ (Eq.~\ref{eq:lambda1}) and (b) full
  posterior probability, $\Lambda_2$ (Eq.~\ref{eq:lambda2}) for the
  empirical networks listed in Table~\ref{tab:data}. The ratio is
  computed so that the preferred model has $\Lambda_{1/2}=1$ and thus
  appears on the top of the figures. The remaining points for each
  dataset correspond to the odds ratio of the remaining models relative
  to the winning one. The solid lines mark a $\Lambda = 10^{-2}$
  confidence threshold. The networks are ordered by increasing number of
  nodes (see table~\ref{tab:data}).\label{fig:model-selection}}
\end{figure*}

\begin{figure}
  \includegraphics[width=.49\columnwidth,trim={.2cm .5cm .45cm .4cm},clip]{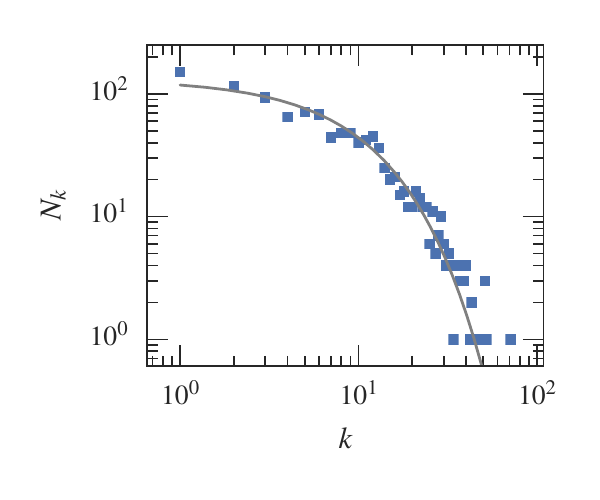}
  \includegraphics[width=.49\columnwidth,trim={.2cm .5cm .45cm .4cm},clip]{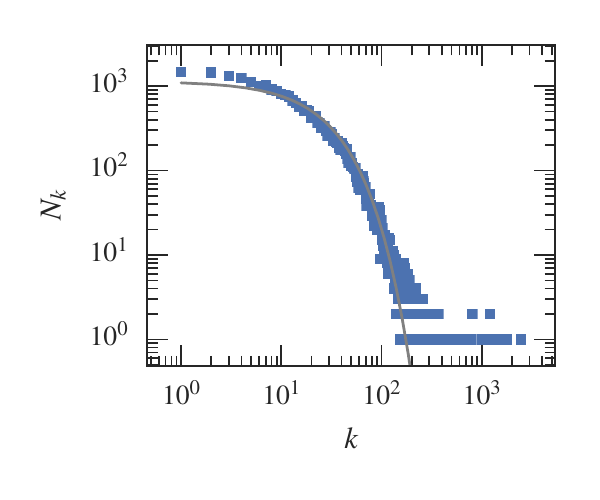}

  \caption{Degree histograms for the Email (left) and arXiv hep-th
  citations (right) networks. In both cases the solid lines show a
  geometric distribution $N_k=Np(1-p)^{k-1}$, with
  $p=1/\avg{k}$.\label{fig:deg-dist}}
\end{figure}

\begin{table}\smaller[2]
\begin{tabular}{l|r|r|r|r|r}
Dataset & $N$ & $\avg{k}$ & $B_1$ & $\avg{B_1}$ & $\sigma_{B_1}$ \\ \hline\hline
Southern women interactions~\cite{davis_deep_2009} &$32$ &$5.6$ &$2$ &$2.4$ &$0.9$ \\
Zachary's karate club~\cite{zachary_information_1977} &$34$ &$4.6$ &$2$ &$2.2$ &$0.5$ \\
Dolphin social network~\cite{lusseau_bottlenose_2003} &$62$ &$5.1$ &$2$ &$2.9$ &$0.5$ \\
Characters in Les Misérables~\cite{knuth_stanford_1993} &$77$ &$6.6$ &$8$ &$8.6$ &$0.7$ \\
American college football~\cite{girvan_community_2002} &$115$ &$10.7$ &$10$ &$10.1$ &$0.3$ \\
Florida food web (wet)~\cite{ulanowicz_network_2005} &$128$ &$32.9$ &$14$ &$14.2$ &$0.4$ \\
Residence hall friendships~\cite{freeman_exploring_1998} &$217$ &$24.6$ &$20$ &$20$ &$0$ \\
\emph{C. elegans} neural network~\cite{white_structure_1986} &$297$ &$15.9$ &$20$ &$13.5$ &$0.5$ \\
Scientific coauthorships~\cite{newman_modularity_2006} &$379$ &$4.8$ &$28$ &$29.6$ &$1.6$ \\
Country-language network~\cite{kunegis_konect:_2013} &$868$ &$2.9$ &$4$ &$10.1$ &$1.9$ \\
Malaria gene similarity~\cite{larremore_network_2013} &$1,104$ &$5.4$ &$56$ &$55.8$ &$1.9$ \\
E-mail~\cite{guimera_self-similar_2003} &$1,133$ &$9.6$ &$28$ &$26.9$ &$0.3$ \\
Political blogs~\cite{adamic_political_2005} &$1,222$ &$31.2$ &$15$ &$15$ &$0$ \\
Scientific coauthorships~\cite{newman_modularity_2006} &$1,589$ &$3.5$ &$48$ &$67.3$ &$3.4$ \\
Protein iteractions (I)~\cite{stelzl_human_2005} &$1,706$ &$7.3$ &$26$ &$40.2$ &$0.6$ \\
Bible names co-ocurrence~\cite{kunegis_konect:_2013} &$1,773$ &$10.3$ &$63$ &$79.1$ &$5.3$ \\
Global airport network~\cite{peixoto_hierarchical_2014} &$3,286$ &$41.6$ &$268$ &$264.6$ &$6.1$ \\
Western states power grid~\cite{watts_collective_1998} &$4,941$ &$2.7$ &$38$ &$37.3$ &$1$ \\
Protein iteractions (II)~\cite{joshi-tope_reactome:_2005} &$6,327$ &$46.6$ &$419$ &$406.4$ &$18.6$ \\
Internet AS~\cite{leskovec_graph_2007} &$6,474$ &$4.3$ &$40$ &$50$ &$7.2$ \\
Advogato user trust~\cite{massa_bowling_2009} &$6,541$ &$15.6$ &$174$ &$80.7$ &$0.6$ \\
Chess games~\cite{kunegis_konect:_2013} &$7,301$ &$17.8$ &$79$ &$79$ &$0$ \\
Dictionary entries~\cite{batagelj_network_2002} &$13,356$ &$18$ &$1,378$ &$1,378.9$ &$2.3$ \\
Cora citations~\cite{subelj_model_2013} &$23,166$ &$7.9$ &$575$ &$575$ &$0.2$ \\
Google+ social network~\cite{leskovec_learning_2012} &$23,628$ &$3.3$ &$46$ &$41.3$ &$2.4$ \\
arXiv hep-th citations~\cite{leskovec_graph_2007} &$27,770$ &$25.4$ &$1,211$ &$1,207.1$ &$4$ \\
Linux source dependency~\cite{kunegis_konect:_2013} &$30,837$ &$13.9$ &$448$ &$384.7$ &$3.1$ \\
PGP web of trust~\cite{richters_trust_2011} &$39,796$ &$15.2$ &$1,350$ &$1,323.2$ &$26.4$ \\
Facebook wall posts~\cite{viswanath_evolution_2009} &$46,952$ &$37.4$ &$6,930$ &$6,794.9$ &$129.9$ \\
Brightkite social network~\cite{cho_friendship_2011} &$58,228$ &$7.4$ &$171$ &$177.4$ &$3.8$ \\
Gnutella hosts~\cite{ripeanu_mapping_2002} &$62,586$ &$4.7$ &$24$ &$24$ &$0$ \\
Youtube group memberships~\cite{mislove_online_2009} &$124,325$ &$4.7$ &$273$ &$266.7$ &$4.7$ \\
\end{tabular}
\caption{Empirical networks used in this work, with their number of nodes
$N$, average degree $\avg{k}=2E/N$, number of groups at the lowest
hierarchical level $B_1$ according to the MDL criterion, and the same
value averaged from the posterior distribution $\avg{B_1}$, as well
as standard deviation of the distribution,
$\sigma_{B_1}$.\label{tab:data}}
\end{table}

We demonstrate the use of our approach on empirical networks (summarized
in Table~\ref{tab:data}), which we also use to compare different model
variations. We begin with a network of political blogs compiled by
Adamic and Glance~\cite{adamic_political_2005} during the 2004 general
election in the USA. In this network nodes are blogs, and an edge exists
between two nodes if one blog cites the other (hence, the network is
directed, and therefore the directed versions of the SBM were used, see
Appendix~\ref{sec:directed}). This network was used in
Ref.~\cite{karrer_stochastic_2011} as an example where the DC-SBM
yielded more meaningful results, since it preferred a partition of the
nodes that was largely compatible with the original categorization done
in Ref.~\cite{adamic_political_2005}, based on the content of the blogs,
into ``liberal'' and ``conservative'' sites. The NDC-SBM, on the other
hand, preferred to divide the nodes only according to degree. However,
in that analysis the number of groups was fixed at $B=2$. Using the
nonparametric approach described here, where the number of groups is
determined from data itself, the results show a less extreme amount of
discrepancy, as seen in Fig.~\ref{fig:polblogs}, which shows the most
likely partition according to each model flavor. In all cases, the
division of the nodes is largely compatible with the accepted one: The
hierarchy branches at the top into the two political factions, and then
proceeds into further sub-divisions inside each group.  However, when
inspecting the lower levels of the hierarchy, we see that the different
variants yield distinct subdivisions inside the two main groups. The
non-degree-corrected version yields the largest number of groups,
followed by the degree corrected one with uniform degree priors, and
finally the version with uniform degree hyperpriors with the smallest
number of groups. In this particular case, the models with smaller
number of groups have also the smallest description length, which seems
to indicate that the division into a larger number of groups are
necessary for the models that are unable to otherwise properly explain
the heterogeneity in the degree sequence. Thus, despite their uniform
agreement with the accepted division, the MDL criterion still confirms
the DC-SBM as a better model for this network.

We now move to a social network between scientists, where an edge exists
if two scientists collaborated on a
paper~\cite{newman_finding_2006}. Here, we compare the results obtained
by employing MDL (i.e. finding the most likely partition) and sampling
many partitions from the posterior distribution, as shown in
Fig.~\ref{fig:netscience}. We observe that while the sampled partitions
share close similarities to the MDL result, there is a noticeable
variance among the individual samples. Fig.~\ref{fig:netscience} also
shows the marginal distribution for the number of groups at the first
three hierarchical levels. For all three model variants, the typical
number of groups is significantly higher that what is obtained for the
optimal partition (due to the low degree variability in this particular
network, it is one of the few that are better modelled by the NDC-SBM,
as seen in Fig.~\ref{fig:model-selection}). This can be understood as an
entropic effect, where the existence of a much larger number of more
complex models with smaller yet comparable likelihood pushes the
posterior distribution towards them. This is a good example of the
bias-variance trade-off mentioned in Sec.~\ref{sec:mdl}, where we see
that the MDL results in a more conservative partition, whereas the full
posterior deposits more collective weight on larger models that are also
more numerous. This seems to indicate that no single partition (and its
associated model) serves as a overwhelmingly better explanation among
those considered --- a symptom that no specific model variant can
perfectly accommodate the network structure, and thus that the SBM is
possibly not a suitable generative model for this data.

This disagreement between MDL and posterior sampling is not universal,
and depends strongly on the network structure. In
Fig.~\ref{fig:B-posterior} we show further results for other networks,
that show a fair amount of diversity in this respect. In many cases the
MDL estimate lies close to the mode of the posterior, indicating a fair
amount of agreement (at least as far as the number of groups is
concerned).

If we compare the different model flavors as outlined in
Sec.~\ref{sec:comparison}, we obtain that most typically the DC-SBM with
uniform degree hyperpriors provides the smallest description length for
a large variety of networks, as shown in
Fig.~\ref{fig:model-selection}a. As expected, the margin by which the
best model is selected increases with the size of the network, as larger
networks typically contain more data. If we compare instead the whole
model class, by summing over all partitions, we obtain largely
consistent (though not identical) outcomes, as seen in
Fig.~\ref{fig:model-selection}b. Exceptions to this include networks
where there is no significant statistical evidence to support the most
complex models --- either due to their small size or
 narrow degree distributions (e.g. Scientific coauthorships, Malaria
 gene similarity and Western states power grid) --- and often the
simpler NDC-SBM is preferred, as well as some networks for which the
DC-SBM with uniform degree priors is preferred instead (E-mail, arXiv
hep-th citations). A closer inspection of these networks reveal that
their global degree distribution is fairly narrow, well approximated by
an exponential distribution, as shown in Fig.~\ref{fig:deg-dist}. Since
this is what is precisely assumed by the uniform degree prior, this
model variation has the advantage in this case. It is worthwhile to
observe that according to both criteria, the preference towards the
DC-SBM over the NDC-SBM is sometimes only attained with the uniform
degree hyperprior. In many cases the NDC-SBM yields a smaller
description length or larger evidence than the degree-corrected variant
with a uniform prior. This means that correcting for arbitrary degree
frequencies --- as opposed to simply the degrees but assuming uniform
frequencies --- can reveal important information on the structure of the
network that would otherwise remain obscured. Nevertheless, our results
seem to validate the intuition behind the DC-SBM as argued in
Ref.~\cite{karrer_stochastic_2011}, that most networks are better
modeled as mixtures of groups with heterogeneous degrees, as opposed to
groups with the homogeneous degrees that are generated by the
NDC-SBM. Importantly, we reach this conclusion aware that the NDC-SBM is
a larger model class with more parameters, since this fact is fully
incorporated in our comparison.

\section{Discussion}\label{sec:conclusion}

The microcanonical approach to the inference of large-scale network
structures offers an opportunity to encode deeper Bayesian hierarchies
into the generative models, which alleviates the underfitting problems
present otherwise, while at the same time enabling the implementation of
efficient inference algorithms with a complexity that is not explicitly
dependent on the number of groups being inferred.

We showed how the degree-corrected SBM can be formulated in a Bayesian
way, via the incorporation of priors for the degree sequence that depend
on the degree distribution, and hence are more capable of decoupling
modular organization from degree regularities. We have again visited the
issue of the maximum number of groups that can be inferred, and
determined that the hierarchical version of the model is significantly
less susceptible to underfitting, by being able to uncover small groups
in very large networks.

We also showed that the microcanonical model is identical to a Bayesian
version of the typical canonical formulation, if we consider only its
shallower version with uniform priors. Hence, the main strength of the
approach presented here lies not in details of the model specification,
but rather in the ease with which higher order Bayesian considerations
can be incorporated.

Throughout the work we have contrasted two approaches to Bayesian
inference, one where we search for the single best network
parametrization (the MDL criterion), and the other where
parametrizations are sampled according to their posterior probability. We
showed that the bias-variance trade-off that these two options represent
can manifest itself in practice, where a lack of quality of fit yields a
disagreement between both approaches. By performing a systematic
analysis of various empirical networks, we observed that the degree of
discrepancy is varied, and itself serves as an indication of the
suitability of the SBM in capturing the network structure.

We argue that the methods proposed here can be useful in the principled
detection of large-scale network structures and in their
interpretation. In particular we believe it can be used as a basis for a
further understanding of the quality of the SBM family of models in
capturing the properties of real networks.

\bibliography{bib}

\appendix

\section{Asymptotic degree distributions sampled from uniform priors and hyperpriors}\label{sec:deg_dist}

We can easily obtain the expected degree distribution when using the
uniform prior for the degree sequence in Eq.~\ref{eq:k_uniform_prior},
if we relax the ensemble to allow the total number of edges to
fluctuate, with the global constraint being enforced only on average. If
we focus on only one group with $N$ nodes and $E$ half edges on average,
a degree sequence $\bm{k}$ will be sampled with a probability that
maximizes the ensemble entropy constrained by the average number of
edges, obtained via the Lagrangian
\begin{equation}
  \mathcal{F} = -\sum_{\bm{k}}P(\bm{k})\ln P(\bm{k}) -\lambda\left(\sum_{\bm{k}}P(\bm{k})\sum_ik_i - E\right),
\end{equation}
where $\lambda$ is a Lagrange multiplier that enforces the
constraint. Obtaining the saddle point $\{\partial \mathcal{F}/\partial
P(\bm{k})=0, \partial \mathcal{F}/\partial\lambda=0\}$ yields the usual
canonical ensemble
\begin{equation}
  P(\bm{k}) = \frac{e^{-\lambda\sum_ik_i}}{Z}.
\end{equation}
The normalization constant is called the partition function, and is
given by
\begin{equation}
  Z = \sum_{\bm{k}}e^{-\lambda\sum_ik_i} = \left(1-e^{-\lambda}\right)^{-N},
\end{equation}
with $\lambda = \ln(1+N/E)$ obtained by enforcing the constraint
$E=\sum_ik_i=-\partial\ln Z/\partial\lambda$. From the above, we obtain
immediately that the probability of a given node $i$ having a degree $k$
is
\begin{equation}
  P(k_i=k) = e^{-\lambda k} \frac{e^{-\lambda\sum_{j\ne i}k_j}}{Z} = (1-e^{-\lambda})e^{-\lambda k}.
\end{equation}
This is a geometric distribution, more commonly parametrized as
\begin{equation}
  P(k) = (1-p)p^k,
\end{equation}
with an average $\avg{k}=(1-p)/p=E/N$. This canonical ensemble is not
identical to the microcanonical one used in the main text, but will
approach it asymptotically in the the thermodynamic limit, i.e. when the
number of nodes and edges become sufficiently large.

We can use the same approach to obtain the expected degree distribution
generated from the uniform hyperprior of Eq.~\ref{eq:k_dist_prior},
which is somewhat more involved, but it is still quite feasible. We want
to consider the ensemble of non-negative integer counts $\{n_k\}$,
subject to a normalization constraint $\sum_{k=0}^{\infty}n_k = N$ and a
fixed average $\sum_{k=0}^{\infty}kn_k = E$. Following the same
maximum-entropy ansatz as above yields a partition function for this
ensemble given by
\begin{equation}
  Z =\sum_{\{n_k\}}e^{-\lambda\sum_kn_k - \mu \sum_kkn_k}=\prod_kZ_k,
\end{equation}
where $\lambda$ and $\mu$ are Lagrange multipliers that keep the
constraints in place, and with
\begin{equation}
  Z_k = \frac{1}{1-\exp(-\lambda - \mu k)}.
\end{equation}
The expected degree counts are given by
\begin{equation}
  \left<n_k\right> = -\frac{\partial \ln Z_k}{\partial\lambda}
                   = \frac{1}{\exp(\lambda + \mu k) - 1},
\end{equation}
which is the Bose-Einstein distribution.  The parameters $\lambda$ and
$\mu$ are determined via the imposed constraints,
\begin{align}
  \sum_{k=0}^{\infty}\frac{1}{\exp(\lambda + \mu k) - 1} = N, \\
  \sum_{k=0}^{\infty}\frac{k}{\exp(\lambda + \mu k) - 1} = E.
\end{align}
For sufficiently large $E$ and $N$, the sums may be approximated by
integrals, and using the polylogarithm function, $\operatorname{Li}_s(z)
= \Gamma(s)^{-1} \int_0^\infty [t^{s-1}/(e^t/z-1)]\dd t$, we
have
\begin{align}
  \int_0^\infty \frac{\dd k}{\exp(\lambda + \mu k) - 1} = \frac{\operatorname{Li}_1(e^{-\lambda})}{\mu} = N, \label{eq:lambda} \\
  \int_0^\infty \frac{k\,\dd k}{\exp(\lambda + \mu k) - 1} = \frac{\operatorname{Li}_2(e^{-\lambda})}{\mu^2} = E. \label{eq:mu}
\end{align}
Eq.~\ref{eq:lambda} can be solved for $\lambda$ as
$e^{-\lambda}=1-\exp(-N/\mu)$, but the same cannot be done for
Eq.~\ref{eq:mu} in closed form. However, for $N \gg \mu$, we have
$\lambda \to 0$, and hence $\mu\approx\sqrt{\operatorname{Li}_2(1) / E}
= \sqrt{\zeta(2) / E}$, with $\zeta(s)$ being the Riemann zeta
function. This yields the asymptotic distribution,
\begin{equation}\label{eq:bose-einstein}
  \left<n_k\right>\approx \frac{1}{\exp\left(k\sqrt{\zeta(2)/E}\right) - 1}.
\end{equation}
Its variance can be obtained from the second moment,
\begin{equation}
  N\avg{k^2} = \int_0^\infty \frac{k^2\,\dd k}{\exp(\lambda + \mu k) - 1} = \frac{\operatorname{Li}_3(e^{-\lambda})}{2\mu^3},
\end{equation}
which leads to
\begin{equation}
  \avg{k^2} = \frac{\zeta(3)}{2}\left(\frac{\avg{k}}{\zeta(2)}\right)^{3/2}\sqrt{N},
\end{equation}
which diverges in the limit $N\gg 1$. For degrees $k \ll \sqrt{E}$, we
have $\exp(k\sqrt{\zeta(2)/E}) \approx 1 + k\sqrt{\zeta(2)/E}$, and
hence the expected distribution of Eq.~\ref{eq:bose-einstein} will
follow a power law $1/k$ for small arguments, with an exponential
cut-off for larger arguments,
\begin{equation}
  \left<n_k\right> \approx
  \begin{cases}
    \sqrt{E/\zeta(2)}/k & \text{ for } k \ll \sqrt{E},\\
    \exp(-k\sqrt{\zeta(2)/E}) & \text{ for } k \gg \sqrt{E}.\\
  \end{cases}
\end{equation}
Distributions of the form $1/k$ are often attributed to non-equilibrium
processes or critical behavior, but as the above shows, they can also
come from maximum-entropy ensembles with simple constraints. This is
tantamount to saying that most discrete distributions with a fixed
average tend to have the above asymptotic form, and therefore no
mechanism other than randomly choosing between them is necessary to
explain this property.

\section{Directed networks}\label{sec:directed}

Although in the main text we focused on undirected networks, directed
model variants are easy to obtain, as we summarize here. For the
directed DC-SBM we have the model likelihood
\begin{equation}
  P(\bm{A}|\bm{k},\bm{e},\bm{b}) = \frac{\prod_ik^+_i!k^-_i!\prod_{rs}e_{rs}!}{\prod_re^+_r!e^-_r!\prod_{ij}A_{ij}!},
\end{equation}
with $k^+_i=\sum_jA_{ji}$, $k^-_i=\sum_jA_{ij}$, $e^+_r=\sum_se_{sr}$,
$e^-_r=\sum_se_{rs}$.  For the hierarchical prior of edge counts, we
have to treat the multigraphs as directed,
\begin{equation}
  P(\bm{e}_l|\bm{e}_{l+1},\bm{b}_l) = \prod_{rs}\multiset{n_r^ln^l_s}{e_{rs}^{l+1}}^{-1}.
\end{equation}
The uniform degree prior is the product of two priors, for the in- and
out-degree sequences,
\begin{equation}
  P(\bm{k}|\bm{e},\bm{b}) = \prod_r\multiset{n_r}{e_r^+}^{-1}\multiset{n_r}{e_r^-}^{-1}.
\end{equation}
Analogously for the conditioned degree prior we need to account for the
joint (in, out)-degree distribution,
\begin{equation}
  P(\bm{k}|\bm{\eta}) = \prod_r \frac{\prod_{k^+,k^-}\eta_{k^+,k^-}^r!}{n_r!}
\end{equation}
and an uniform hyperprior
\begin{equation}
  P(\bm{\eta}|\bm{e},\bm{b}) = \prod_r q(e_r^+, n_r)^{-1}q(e_r^-, n_r)^{-1}.
\end{equation}
The NDC-SBM is also entirely analogous, corresponding to a degree
probability
\begin{equation}
  P(\bm{k}|\bm{e},\bm{b}) = \prod_r\frac{e_r^+!}{n_r^{e_r+}\prod_{i\in r}k_i^+!}\prod_r\frac{e_r^-!}{n_r^{e_r-}\prod_{i\in r}k_i^-!},
\end{equation}
which yields the model likelihood
\begin{equation}
  P(\bm{A}|\bm{e},\bm{b}) = \frac{\prod_{rs}e_{rs}!}{\prod_rn_r^{e_r^+}n_r^{e_r^-}}\times
  \frac{1}{\prod_{ij}A_{ij}!}.
\end{equation}

\end{document}